\documentclass[acmlarge,screen]{acmart}
\usepackage{amsmath}
\usepackage{tikz}
\usepackage{subcaption}
\usepackage{pifont}
\usepackage{siunitx}
\usepackage{lipsum}
\usepackage{colortbl}
\usepackage{enumitem}
\usepackage{xcolor}
\usepackage{makecell}
\usepackage{algorithm}
\usepackage{algpseudocode}
\usepackage{graphicx}
\usepackage{booktabs}
\usepackage{multirow}
\usepackage{soul}
\usepackage{chngcntr}

\usepackage{array,adjustbox,tabularx}
\newcolumntype{Y}{>{\centering\arraybackslash}X}
\definecolor{myblue}{RGB}{79,173,234}

\AtBeginDocument{%
  }
\setcopyright{cc}
\setcctype{by}
\copyrightyear{2026}
\acmYear{2026}
\acmDOI{10.1145/3810205}
\acmJournal{IMWUT}
\acmYear{2026} \acmVolume{10} \acmNumber{2} \acmArticle{74}
\acmMonth{6}

\newcommand{\dimens}[2]{$#1 \times #2$}
\newcommand{\squishlist}
{\begin{itemize}[itemsep=1pt,parsep=2pt,topsep=3pt,partopsep=0pt,leftmargin=0em, itemindent=1em,labelwidth=1em,labelsep=0.5em]}
\newcommand{\squishend}{\end{itemize}}
\newcommand{\squishenum}{\begin{enumerate}[itemsep=1pt,parsep=2pt,topsep=3pt,partopsep=0pt,leftmargin=0em, itemindent=1.5em,labelwidth=1em,labelsep=0.5em]}
\newcommand{\squishsubenum}{\begin{enumerate}[itemsep=1pt,parsep=2pt,topsep=0pt,partopsep=0pt,leftmargin=0em,listparindent=1.5em,labelwidth=1em,labelsep=0.5em]}
\newcommand{\squishenumend}{\end{enumerate}}
\newcommand{\uL}{\si{\micro\liter}}
\newcommand{\cmark}{\textcolor{green!60!black}{\ding{51}}}  % check mark
\newcommand{\xmark}{\textcolor{red}{\ding{55}}}             % cross mark
\newcommand{\red}[1]{\textcolor{black}{#1}}
\newcommand{\rednew}[1]{\textcolor{black}{#1}}

\newcommand{\sysname}{\textit{DropleX}}
\makeatletter

\author{Siqi Zhang}
\orcid{0009-0001-1765-1331}% TODO: replace with Siqi Zhang's actual ORCID
\affiliation{
  \institution{Carnegie Mellon University}
  \city{Pittsburgh}
  \state{Pennsylvania}
  \country{USA}
}
\email{siqiz2@andrew.cmu.edu}

\author{Mayank Goel}
\orcid{0000-0003-1237-7545}% TODO: replace with Mayank Goel's actual ORCID
\affiliation{
  \institution{Carnegie Mellon University}
  \city{Pittsburgh}
  \state{Pennsylvania}
  \country{USA}
}
\email{mayankgoel@cmu.edu}

\author{Justin Chan}
\orcid{0000-0002-1471-6187}% TODO: replace with Justin Chan's actual ORCID
\affiliation{
  \institution{Carnegie Mellon University}
  \city{Pittsburgh}
  \state{Pennsylvania}
  \country{USA}
}
\email{justinchan@cmu.edu}

\begin{document}

\begin{CCSXML}
<ccs2012>
   <concept>
       <concept_id>10003120.10003138.10003140</concept_id>
       <concept_desc>Human-centered computing~Ubiquitous and mobile computing systems and tools</concept_desc>
       <concept_significance>500</concept_significance>
       </concept>
 </ccs2012>
\end{CCSXML}

\ccsdesc[500]{Human-centered computing~Ubiquitous and mobile computing systems and tools}

% \settopmatter{printacmref=false,printfolios=false}
\keywords{liquid testing, capacitive sensing}

% \title{{\sysname}: Droplet sensing using tablet touchscreens}
\title{{\sysname}: Liquid Sensing on Tablet Touchscreens}

\begin{teaserfigure}
\centering
\includegraphics[keepaspectratio, width=0.8\linewidth]{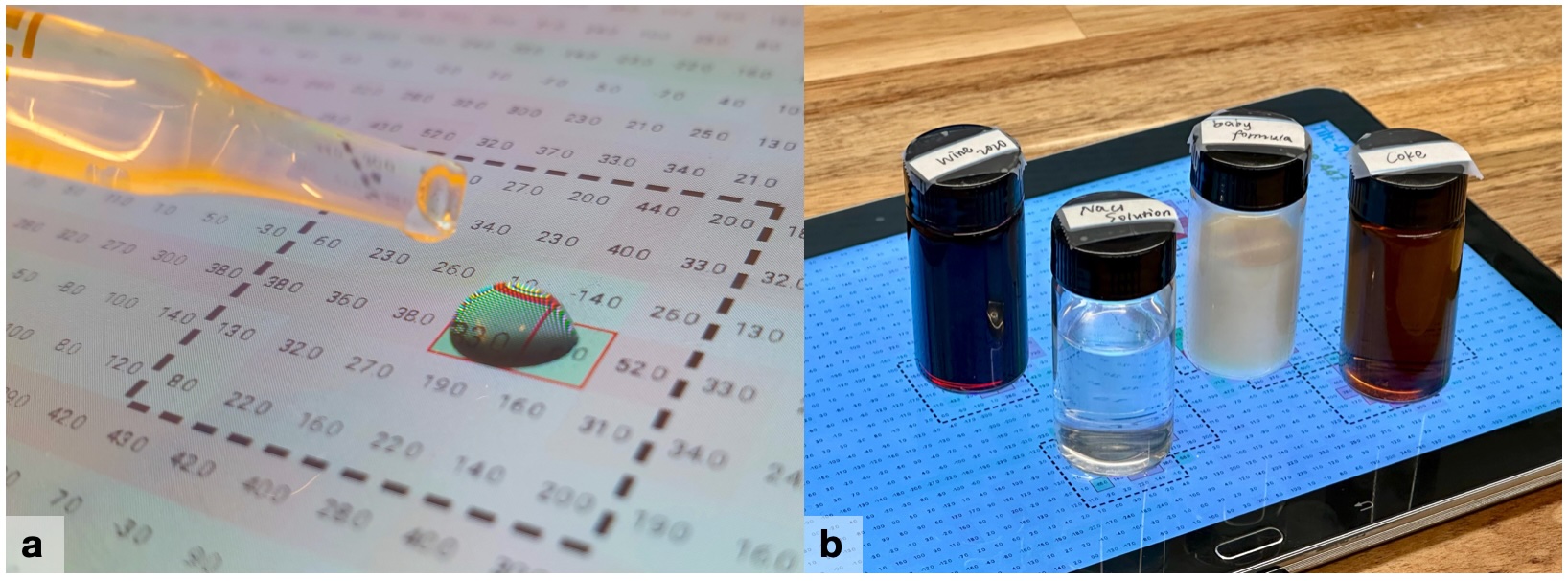}
  \vspace{0em}
  % \caption{{\sysname} explores three key capabilities of liquid sensing on capacitive screens on tablets: {\bf (a)} droplet-scale sensing {\bf (b)} multi-sample sensing {\bf (c)} through-container sensing.}
  % \caption{{\sysname} enables liquid sensing on capacitive screens on tablets for {\bf (a)} microliter-scale sensing {\bf (b)} concurrent multi-sample sensing {\bf (c)} through-container sensing. These capabilities have broad applications in detecting adulteration and contamination in everyday liquids, flagging trace biomolecular chemicals, detecting drink spiking as well as being a ``lab-on-a-pad'' platform for accessible scientific education at scale.}  
  \caption{{\sysname} enables liquid sensing on tablet capacitive touchscreens for {\bf (a)} microliter-scale sensing and {\bf (b)} concurrent multi-sample through-container sensing. The figure shows our custom real-time app sensing samples placed directly on the display and through-containers.}
  \label{fig:fig1}
\end{teaserfigure}

\begin{abstract}
We present {\sysname}, the first system that enables liquid sensing using the capacitive touchscreen of commodity tablets. {\sysname} detects microliter-scale liquid samples, and performs non-invasive, through-container measurements \red{for liquid analysis}. These capabilities are made possible by a physics-informed mechanism that disables the touchscreen's built-in adaptive filters, originally designed to reject the effects of liquid drops such as rain, without any hardware modifications. We model the touchscreen's sensing capabilities, limits, and non-idealities to inform the design of a signal processing and learning-based pipeline for liquid sensing. \rednew{Under controlled laboratory conditions, our }system achieves 89--99\% accuracy in detecting microliter-scale adulteration in soda, wine, and milk, 94--96\% accuracy in threshold detection of trace chemical concentrations, and 86--96\% accuracy in through-container adulterant detection. \red{These exploratory results demonstrate the potential of repurposing commodity touchscreens as a liquid characterization platform for laboratory settings, food and beverage testing, and chemical analysis applications.}
\end{abstract}

\maketitle

\section{Introduction}
% While touchscreens have conventionally been designed to reject the effects of liquids such as raindrops, treating it as unwanted noise, this paper takes the opposite approach and investigates if instead commodity capacitive touchscreens can be repurposed to perform liquid sensing. Given the predominance of capacitive touchscreens on modern computing devices, the ability to perform liquid sensing on touchscreens could transform tablets into a portable ``lab-on-a-pad'' as an accessible means of adulteration and concentration detection in everyday settings, resource-constrained environments and classrooms.

Capacitive touchscreens on smart devices such as tablets have conventionally been designed to reject the effects of liquids such as raindrops, treating them as unwanted noise. In this work, we take the opposite approach and investigate whether commodity capacitive touchscreens can instead be repurposed to perform liquid sensing. \red{By analyzing how microliter-scale liquid samples and larger liquid volumes in containers interact with capacitive touchscreens, we explore the feasibility of using these devices for liquid characterization in controlled settings. }This approach could enable tablets to serve as accessible liquid analysis platforms for laboratory use, food and beverage quality testing, and chemical characterization in educational or resource-constrained environments.

Conventional methods~\cite{conventional,conventional2,conventional3} for performing capacitance-based liquid sensing typically rely on a dedicated device with large probes designed to be immersed into open cup solutions to a minimum depth, and are typically limited to measuring a single sample at a time. While there is rich prior work~\cite{yeo2016radarcat,rahman2016nutrilyzer,wang2017tagscan,dhekne2018liquid,yue2019liquid,ha2020food,yun2024powdew,chan2022testing,chan2022micro} in the mobile and wireless community on liquid based sensing using radar, RFID, vibration motors and cameras, including work leveraging an external standalone capacitive touchscreen, they are largely focused on sensing larger quantities of liquid, require some form of dedicated, specialized hardware and non-trivial hardware modification to commodity devices, or do not demonstrate the ability to classify between different liquids. 

We introduce {\sysname}, the first system to enable liquid sensing on the integrated capacitive touchscreen of commodity tablets. Our system is capable of detecting microliter-scale liquid samples and performing non-invasive through-container measurements. It is able to conduct multiple tests concurrently across the screen, allowing simultaneous characterization of different samples.

However, repurposing the tablet touchscreen to perform liquid sensing is not straightforward and requires tackling several key challenges. \textit{First}, commercial touchscreens have built-in adaptive filters that continuously recalibrate capacitance readings to a baseline to reject transient effects such as those caused by liquid drops. As a result, any liquid or container placed on the screen is only registered for a few seconds before the readings fades. To address this, we develop a physics-informed mechanism that can disable the filter temporarily for the measurement without any hardware modifications.

Specifically, we observe that the adaptive filter is turned off in the presence of a finger touch event. This is because recalibrating during contact would cause the system to disregard the touch input, and make touch interaction impossible. Our idea is to mimic a permanent touch event, by having the user deposit and draw up a small ``priming'' drop of water onto the screen which leaves behind a thin conductive film. Unlike a bulk drop of water that confines the electric field to the screen, the thin film acts primarily as a conductor and causes part of the electric field to be shunted to parasitic ground and is effectively recognized as a finger. This causes the adaptive filter to be disabled and allows for reliable measurements of liquids placed on the screen. 

This deposition and removal process can be performed using common applicators including eye droppers, and can also be performed without any special equipment. The user can simply dip their finger into a cup of water, allowing a small pendant drop to adhere to the skin, and then touch the screen to deposit it. The drop can then be drawn up by lightly touching it with a tissue, which wicks it away through capillary action.

\textit{Second}, the measured capacitance values for different liquids or concentrations can overlap, making it difficult to distinguish using a single scalar metric. To overcome this, we instead leverage the spatial spreading patterns of the liquid, as well as the fringing fields from the electric field lines that extend beyond the drop boundary, and leverage a learning-based approach to capture pertinent features that can accurately characterize the drop.

\textit{Third}, as touchscreens are not designed for precise measurements of capacitance for liquid sensing, the electrode grid exhibits spatially varying sensitivity and time-varying noise across the screen. To address this challenge, we develop a one-time, multi-point calibration method to generate a sensitivity compensation map over the screen, and apply temporal averaging to reduce noise.

\noindent Our contributions are as follows:
\squishlist
\item We present {\sysname}, the first system that enables liquid sensing on commodity tablet touchscreens without any hardware modifications. \red{Under controlled laboratory conditions, we} demonstrate adulteration detection accuracies of 89--99\% for soda, wine, and milk samples, detection of trace amounts of biochemicals including DNA, NaCl, ethanol with accuracies of 94--96\%, and through-container sensing with accuracies of 86--96\% across multiple containers. \rednew{These results should be interpreted as a proof-of-concept and a system for use in uncontrolled real-world settings remains a direction for future work.}
\item We run characterization and benchmark experiments across different sample volumes, temperatures, as well as the effect of interfacial layers such as thin-films and different containers on the capacitance readings.
\item We develop a physics-informed model that relates the device's measurements to physical capacitance readings.
\item We develop an end-to-end software infrastructure, including utilities to read the capacitance heatmaps, a real-time Android tablet app and desktop app to visualize the readings. We plan to open-source the code and tutorial for enabling this capability on tablets as well as our dataset upon publication as a key technical primitive that the community can build on to enable future sensing systems on touchscreens.
\squishend

\red{As the proposed approach involves measurements of capacitance heatmaps already accessible to technology manufacturers in software, it can in principle be deployed through a software or firmware update to commodity devices, making it suitable as a liquid characterization tool for controlled testing environments.}
\section{Related work}

\begin{table*}[h]
\centering
\small
\renewcommand{\arraystretch}{1.2} % more row padding
\begin{tabularx}{\textwidth}{|l|Y|Y|Y|Y|Y|}
\hline
\textbf{Reference} & \textbf{Technology} & \textbf{Quantity measured} & \textbf{Ubiquitous smart device} & \textbf{No hardware additions or modifications needed} & \textbf{Distinguishes between concentrations} \\ \hline
\multicolumn{6}{|c|}{\textbf{Macroscopic-scale methods (\(>1~\mathrm{mL}\))}} \\ \hline
RadarCat (2016)~\cite{yeo2016radarcat}&	Radar&	Reflected signal&	\xmark&	N/A&	\xmark\\ \hline
Nutrilyzer (2016)~\cite{rahman2016nutrilyzer}& 	Photoacoustic effect&	Photoacoustic spectra&	\cmark&	\xmark&	\cmark \\ \hline
TagScan (2017)~\cite{wang2017tagscan}&	RFID&	RSS, phase change&	\xmark&	N/A&	\cmark\\ \hline
LiquiID (2018)~\cite{dhekne2018liquid} & Ultra-wideband radios & Permittivity & \xmark & N/A & N/A \\ \hline
CapCam (2019)~\cite{yue2019liquid}&	Vibration motor, camera, ripples&	Surface tension&	\cmark&	\cmark&	\cmark\\ \hline
RF-EATS (2020)~\cite{ha2020food}&	RFID&	Amplitude and phase change&	\xmark&	N/A&	\cmark\\ \hline
% PowDew (2024)~\cite{yun2024powdew}&	Smartphone display and camera&	Droplet motion, spreading, and penetration&	\cmark&	\cmark&	\cmark\\ \hline
\multicolumn{6}{|c|}{\textbf{Microliter-scale methods (\(50~\uL~\text{to}~1~\mathrm{mL}\))}} \\ \hline
\makecell[l]{Prior touchscreen\\systems (2016--2021)\\~\cite{horstmann2021capacitive,won2012ga,won2016new,lee2018label}}&	Capacitive sensing&	Conductivity&	\xmark&	\xmark&	\xmark\\ \hline
Chan et al. (2022)~\cite{chan2022testing}&	LiDAR, laser speckle, camera&	Viscosity&	\cmark&	\xmark&	\cmark\\ \hline
Chan et al. (2022)~\cite{chan2022micro}&	Vibration motor and camera&	Viscosity&	\cmark&	\cmark&	N/A\\ \hline
\rowcolor{gray!15}
{\bf {\sysname} (ours)}&	{\bf Capacitive sensing}&	{\bf Capacitance}&	\cmark&	\cmark&	\cmark\\ \hline
\end{tabularx}
\caption{{\bf Comparison of liquid sensing systems.} Representative works that focus on mobile and wireless systems for liquid sensing. Prior works using touchscreens for microliter-scale liquid sensing use external standalone touchscreens which are not integrated into smart devices. Our work addresses the practical challenges associated with microliter-scale sensing and classification on tablet touchscreens.}
\vspace{-2em}
\label{tab:rw}
\end{table*}

Prior work can be broadly divided into three different domains (Table~\ref{tab:rw}).

\noindent {\bf Touchscreens for liquid sensing.} Several works~\cite{horstmann2021capacitive,won2012ga,won2016new,lee2018label} have proposed the use of capacitive touchscreens to measure conductivity properties of different electrolytes, polymers, and biomolecules including DNA such as of the H1N1 virus, Chlamydia trachomatis. However, they suffer from several key limitations: {\bf (1)} they use an external standalone display which is not part of a smart device, and do not face or address the practical challenge of an adaptive filter {\bf (2)} they uses an electrode array pitch (separation) that is 2 to 3.5 times smaller than what is found in conventional tablet touchscreens, {\bf (3)} it coats the touchscreen surface with a protective polymer of parylene C, which is impractical for everyday use. Collectively, these constraints fail to capture the practical realities of performing liquid sensing on commodity touchscreens integrated into tablets, and therefore do not encounter or address the range of real-world challenges our work considers. Moreover, prior efforts do not attempt to classify between different liquid types or concentrations, and their reported capacitance values often exhibit significant overlap across samples.

Our work differs in three important aspects: {\bf (1)} It performs liquid sensing directly on unmodified tablet touchscreens, and overcomes the challenges of hardware filters and other non-idealities to make this possible. {\bf (2)} It goes beyond reporting raw measured capacitance values and develops a pipeline to perform liquid classification for different classes of liquids. {\bf (3)} Importantly, we demonstrate through-container sensing scenarios for a variety of containers, which enables use-cases in everyday settings outside lab testing of liquids.

\noindent {\bf Touchscreens for novel interaction techniques.} Beyond liquid sensing, there has also been interest in leveraging smart device touchscreens for enabling novel and tangible interaction capabilities such as by detecting ungrounded conductive objects on the screen~\cite{voelker2013pucs,voelker2015percs,rekimoto2002smartskin}, applying super-resolution to detect the visual signature of the object through motion~\cite{mayer2021super}, estimating finger contact angle~\cite{xiao2015estimating} detecting between fingertips, raindrops and water smear~\cite{tung2018raincheck}, and differentiating between different user's handprints for authentication~\cite{guo2015capauth}. 

A key challenge with enabling this is that commercial touchscreens specifically implement an adaptive filter to detect fingers which are electrically grounded, and reject other objects. Various works~\cite{jansen2012tangible,kratz2011capwidgets,yu2011tuic,yu2011clip} have presented alternative approaches to disabling this filter such as having the body capacitance of the user couple to the touchscreen such as through a conductive element connected to the screen~\cite{wiethoff2012sketch}, or using a conductive element connected to a relatively grounded object such as the battery ground connector of the tablet or a second area of the display as ground~\cite{voelker2013pucs}. The challenge with this approach is that it uses dedicated custom conductive markers, and the grounding is dependent on the geometry of the device. In contrast, our approach leverages the existing benign liquids such as water which is ubiquitous as a means of tricking the phone into thinking there is a grounded object in contact with the screen and overriding the adaptive filter. We note further that none of these systems are designed for liquid sensing and are focused around detecting tangible objects or widgets on the screen. Our work is focused on sensing the properties of micro-liter-scale samples of liquid based on mutual capacitance readings of the touchscreen.

\noindent {\bf Liquid-based sensing.} There is rich prior work in the space of liquid sensing using mobile and wireless systems ranging from using radar-based approaches including UWB~\cite{dhekne2018liquid}, RFID~\cite{wang2017tagscan,xie2019tagtag, ha2020food,ha2018learning}, mmWave radar~\cite{yeo2016radarcat}, photoacoustic-based~\cite{rahman2016nutrilyzer} approaches for liquid contaminant detection. These systems offer complementary designs to the touchscreen-based approach and are designed to operate on liquid samples that are at least 10~mL which is three orders of magnitude larger than what touchscreen-based approaches can measure. These systems are also focused on using custom, dedicated hardware systems which can be challenging to scale outside laboratory settings in particular to resource-constrained environments.

% PowDew~\cite{yun2024powdew} is a complementary work focused on detection of counterfeit powdered food products by analyzing the interaction of a water droplet on powder formula and characterizing its spreading and penetration in the powder using a checkboard pattern projected from the display onto the powder and using the smartphone camera to capture the droplet interactions. This work is complementary our touchscreen system in that it focuses on using unmodified smart devices and a drop of liquid, but is focused around the contamination of powdered food which is a distinct and separate challenge.

Prior work using mobile devices for liquid sensing includes CapCam~\cite{yue2019liquid} which leverages the vibration motor on the smartphone placed on top of a cup to vibrate the liquid and measure the wavelength of the capillary waves and compute surface tension. The constraint of this approach is that the sample needs to be placed in a cup that is filled to a depth of 25 to 45~mm so that the focal length of the camera lens matches the depth of the liquid, which excludes it from practical use of analyzing smaller liquid samples. Chan et al.~\cite{chan2022micro} measured blood clotting time on smartphones for 10~\uL~samples of blood and plasma using a vibration motor, camera, and a tiny copper particle that needs to be deposited into each sample. A similar related work~\cite{chan2022testing} also performed blood clot testing using the LiDAR sensor and laser speckle to measure the viscosity of 10~\uL~samples of blood, milk, and other household liquids. However, these systems require custom hardware components or modifications: specialized copper particles, or smartphone cameras with the near-infrared filter removed to analyze the LiDAR-based laser speckle patterns. They are also focused on sensing different liquid properties: surface tension and viscosity, which rely on mechanical sensing as opposed to electrical sensing, which is what our work focuses on.

Our work is complementary to these approaches in spirit in that it is focused on leveraging commodity smart devices for liquid sensing, but it does not require any hardware modification or special hardware components to perform liquid sensing. 
\section{Enabling liquid sensing on touchscreens}

\begin{figure}[t]
\centering
{\includegraphics[keepaspectratio, width=\linewidth]
    {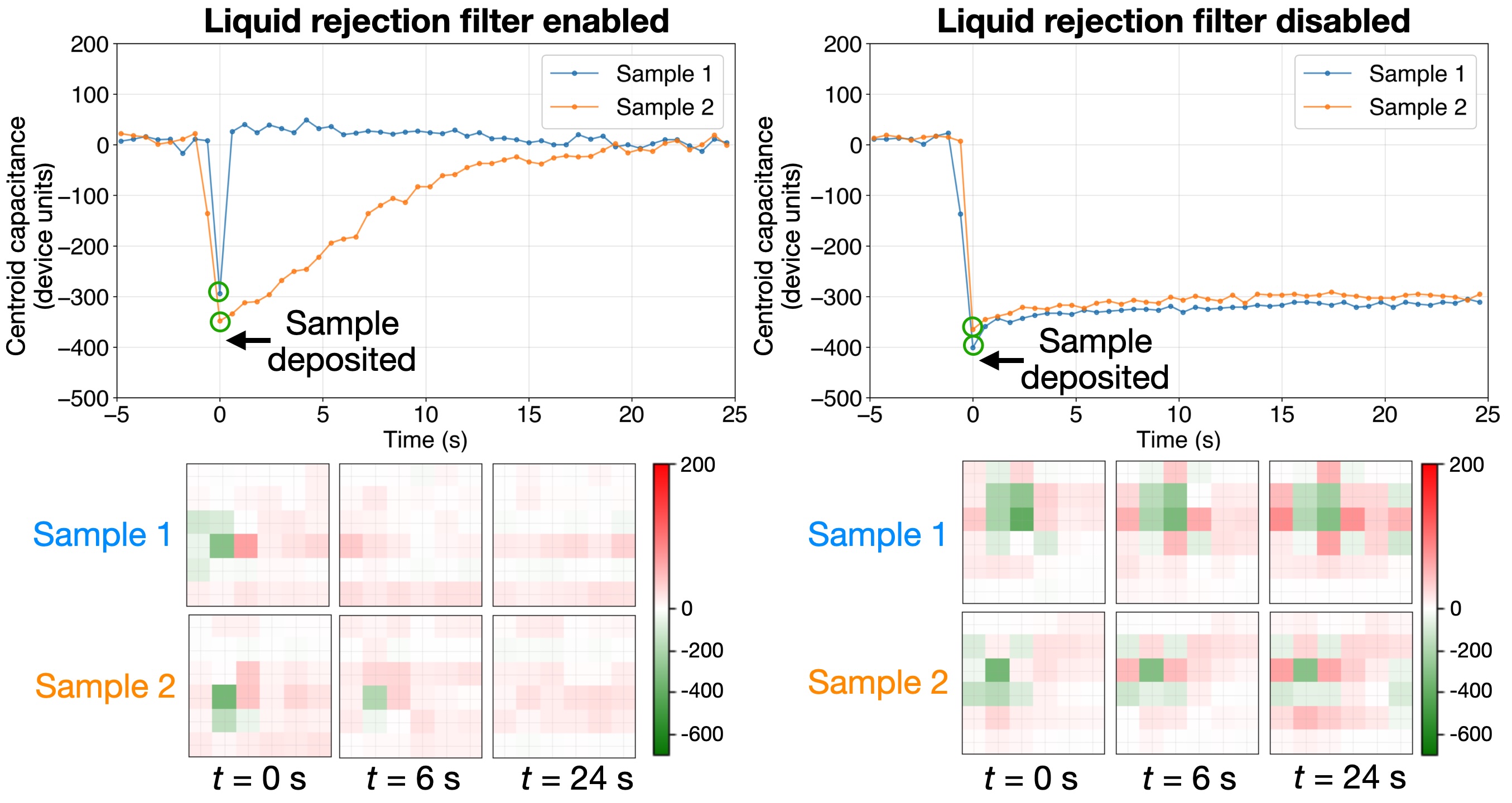}}
  \caption{{\bf Effect of liquid rejection filter on measured capacitance values.} {\bf (a)} With the filter enabled by default, the touchscreen controller gradually re-adapts the baseline, causing the measured signal to return to its original level over time scales of approximately 1–20~s. {\bf (b)} When the filter is disabled, the sample is continuously registered and does not decay away over time. We note that the device units of capacitance have the inverse sign of the actual capacitance change.}
  % \vspace{-1em}
  \label{fig:reject_challenge}
\end{figure}

{\bf Effect of touchscreen’s liquid rejection filter.} While it may appear straightforward to read the capacitive readings from the tablet to perform liquid sensing, an immediate challenge becomes apparent when depositing a drop of water on the screen: the capacitive readings fade away. Fig.~\ref{fig:reject_challenge}a shows the capacitance measured using our tablet's reported digital units over a period of 25 seconds for two drops of 500~\uL water. The capacitance readings are logged at 1.67~Hz corresponding to a frame measured every 0.6~s, implementation details provided in Sec.~\ref{sec:impl}.

The plot reveals two observations, first there is an initial sharp drop in the capacitance readings (in device units). Second, the measured capacitance reverts back to baseline. For sample 1, it reverts back to baseline quickly, while for sample 2, this change was more gradually. 

The reason is that manufacturers specifically design software heuristics known as \textit{adaptive filters} that actively reject liquids such as rain and other spurious events that change the capacitance readings. Disabling this adaptive filter from constantly recalibrating to a baseline would enable our system to measure the effect of liquid samples on the capacitance values more reliably. 

We observe that the adaptive filter is disabled when a finger, the intended input for touchscreens, is in contact with the display. This is expected behavior, since the touchscreen must maintain a persistent signal for a valid touch.

\begin{figure}[t]
\centering
{\includegraphics[keepaspectratio, width=\linewidth]
    {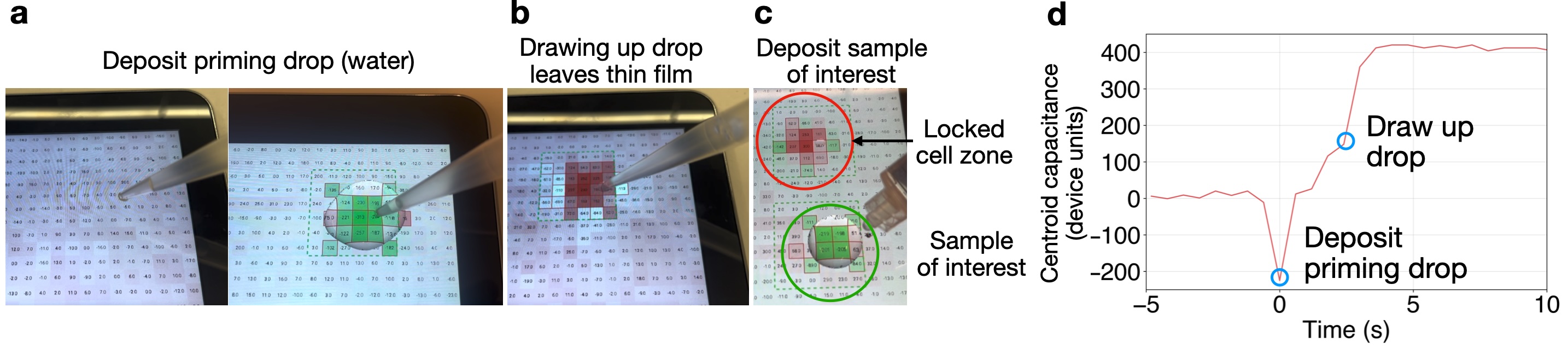}}
  \caption{{\bf Disabling the touchscreen's liquid rejection filter by mimicking a permanent finger touch event.} {\bf (a)} Depositing a priming drop (water sample) on the screen increases the capacitance as it is dominated by the permittivity effect, and the electric field is confined within the sample. {\bf (b)} We disable the filter by mimicking a permanent finger touch event. To do this, we draw up the priming drop which leaves a thin film that causes the electric field to be shunted to parasitic ground, similar to a finger. {\bf (c)} With the adaptive filter disabled, subsequent samples of interest to be measured can be deposited on the screen. {\bf (d)} Shows the capacitance at the centroid of the priming drop when it is initially deposited then drawn up.}
  % \vspace{-1em}
  \label{fig:reject_solution}
\end{figure}

\begin{figure}[t]
\centering
{\includegraphics[keepaspectratio, width=.7\linewidth]
    {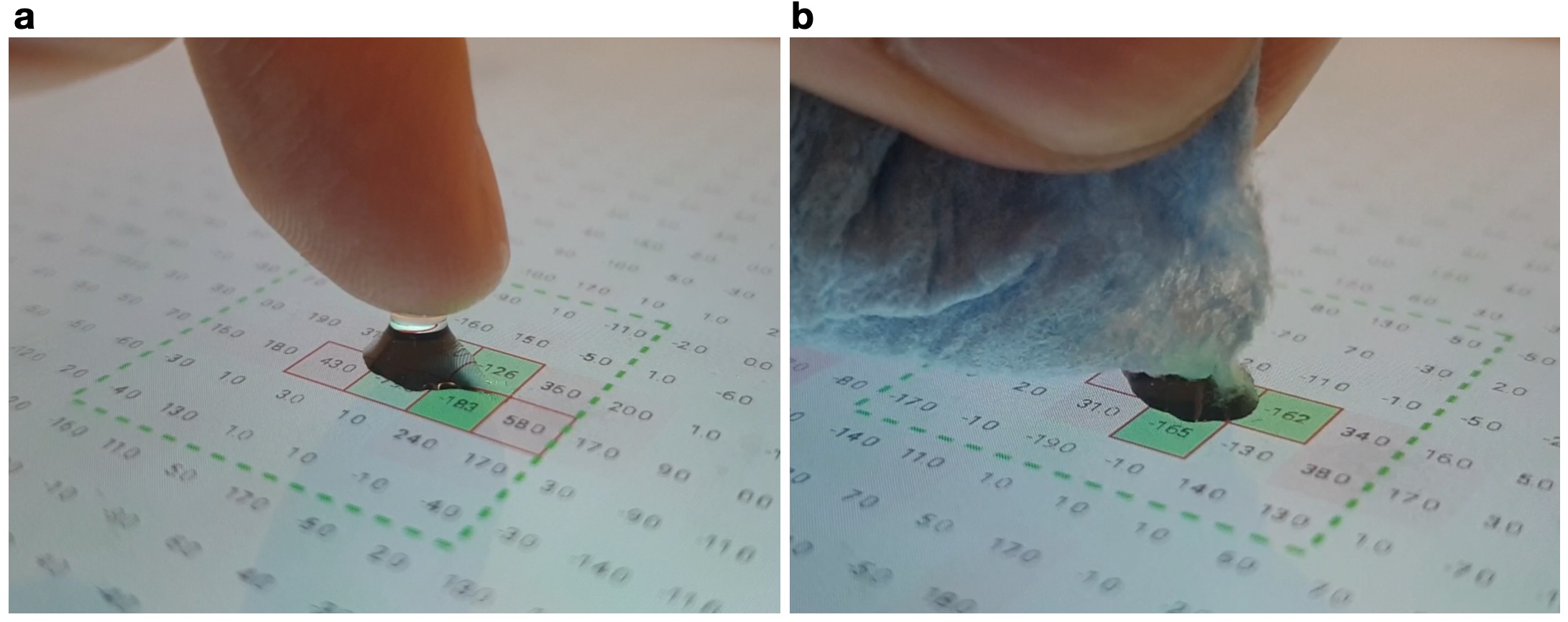}}
  \caption{{\bf Equipment-free deposition and drawing up of priming drop.} {\bf (a)} User deposits pendant drop onto screen, and {\bf (b)} draws up priming drop using tissue paper via capillary wicking.}
  % \vspace{-1em}
  \label{fig:finger}
\end{figure}

\noindent {\bf Mimicking a permanent touch to disable the filter.} It is however impractical to expect the user to constantly place their finger on the screen throughout the measurement. Our key idea is to mimic a permanent touch event to disable the adaptive filter and enable liquid sensing. Our approach is based on our modeling of how a liquid sample affects the touchscreen's capacitive field (Fig.~\ref{fig:bg2}). Specifically, when a liquid is deposited in the screen both its conductivity and permittivity change the capacitive field. The dielectric permittivity term dominates compared to conductance, because a sample of liquid is thick, and in the case of water can be 2~mm in height due to its surface tension, which can be thicker than tablet touchscreens which can have a thickness of 1~mm~\cite{thickness}. Because of this the electric field is largely confined to within the liquid. 

If the liquid height could be reduced to a thin film for the electric field to pass through, the conductivity term would dominate and it would function like a finger. {\textit{Our key observation is that when drawing up the liquid, a thin film remains which causes the electric field to be shunted to parasitic ground, and is effectively recognized by the screen as a ``finger''.}}

To demonstrate this, we deposit a 500~\uL~``priming drop'' of water onto the screen which initially induces a positive change in measured capacitance (negative in device units). After this, upon drawing up the sample using the pipette the capacitance values become negative (positive in device units) and causes a region of pixels to be in a permanently ``stuck'' position, effectively emulating a permanent touch. Fig.~\ref{fig:reject_solution}b shows that depositing this initial sample results in a sharp drop of around -200 device units, and after it is drawn up, it reaches a stable a value of approximately 400 device units. 

% The reason for this effect is that when the water is initially deposited the capacitance reading is dominated by the conductivity effect where charge is drawn away from the screen and towards ground hence reducing the measured capacitance. However, upon removing the liquid, a thin residual film still remains on the screen. Due to its limited thickness, it does not provide a strong ground path and instead acts primarily as a dielectric. Here, the capacitive reading is dominated by the permittivity effect and the residual polarization increases the effective capacitance.

After the adaptive filter is disabled, the liquid sample of interest can be deposited elsewhere on the screen without diminishing. We show in Fig.~\ref{fig:reject_challenge}b that after disabling the adaptive filter, we deposit two samples of tap water as before. This results in a sharp drop in mutual capacitance, after which the sample readings remain stable. 

{\textit{Importantly, we note that a pipette or specialized applicator is not required to deposit or draw up the priming drop.}} As show in Fig.~\ref{fig:finger}, the user can simply dip their finger into water, drop the water on the touchscreen, and then immediately remove it using a regular tissue, which draws the liquid away via capillary wicking.

We note that prior works~\cite{voelker2013pucs} for tangible user interfaces which place ungrounded conductive objects on the screen (e.g. copper tape, coins, keys) have noted this effect and face a similar challenge. These prior works propose including a small conductive clip to ground a point on the screen through capacitive coupling with the chassis of the tablet or as a bridge between electrodes on the screen in order to permanently ground it. However, this requires a custom clip or conducting bridge. Our proposed approach relies on just using tap water which is ubiquitous.
% we find this approach is not as reliable at disabling the filter (we evaluate this in a benchmark in Sec.~\ref{sec:eval}) and

\begin{figure}[h]
\centering
{\includegraphics[keepaspectratio, width=.6\linewidth]
    {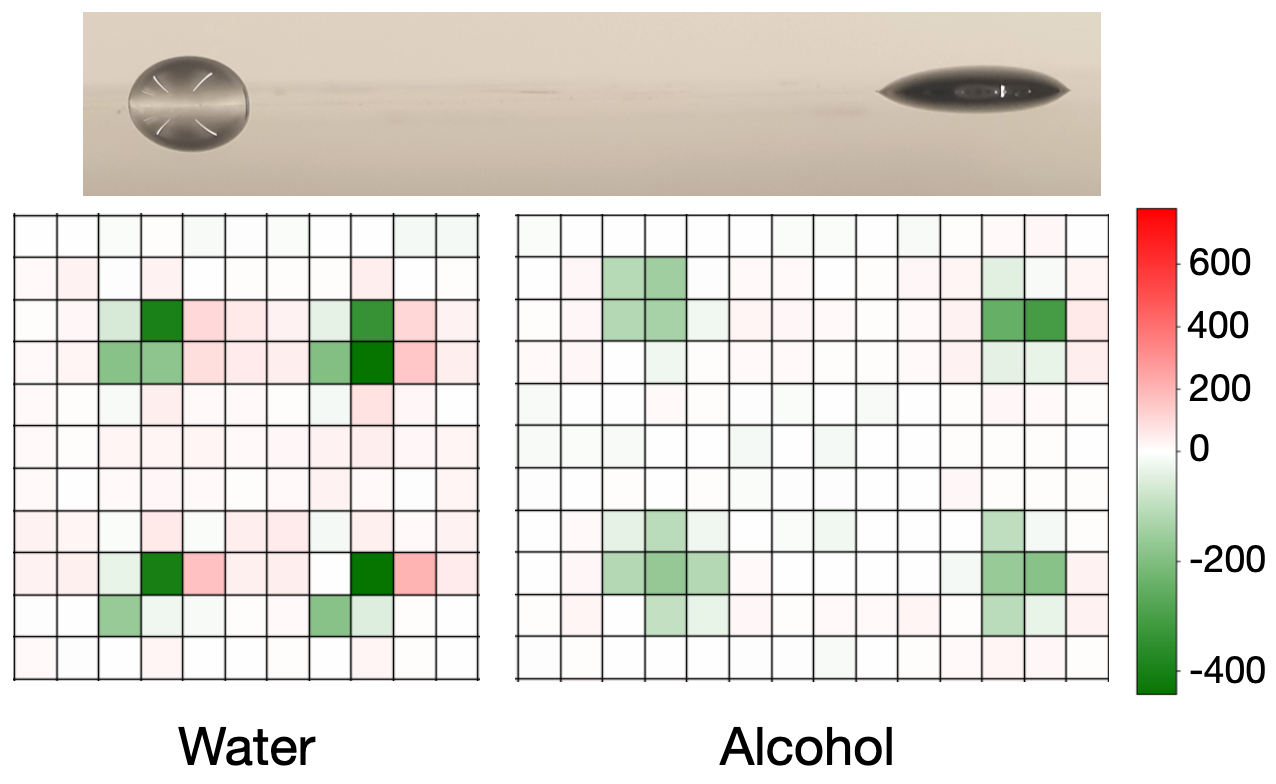}}
  \caption{{\bf Liquids drops measured on tablet touchscreen.}}
  % {\bf (b)} sample temperature, and {\bf (c)} thin-film insulating layer. Note that these are the raw device values without compensation for spatial and temporal variation.}
  \vspace{-1em}
  \label{fig:surface_tension}
\end{figure}

% \begin{figure}
% \centering
% {\includegraphics[keepaspectratio, width=0.8\linewidth]
%     {figs/container/container_stats.png}}
%   \caption{{\bf Through-container liquid positive mean coefficient.}}
%   \label{fig:container_stats}
% \end{figure}

\noindent {\bf Capacitance Heatmap Signatures of Liquids.} We show in Fig.~\ref{fig:surface_tension} an example capacitance heatmap measured after disabling the adaptive filter for 200~\uL~drops of tap water and isopropyl alcohol. The plot shows that different liquids are able to produce different signatures which can be captured by the (1) spatial spreading patterns, (2) capacitance values at the centroid and adjacent cells, (3) fringing patterns around the drop which occur because electric field lines between electrodes extend beyond the droplet boundary and couple to the liquid. We note these fringing patterns are more pronounced in liquids with higher permittivity such as tap water.

\section{Modeling capacitance measurements}
\label{sec:theory}

\begin{figure}[t]
\centering
{\includegraphics[keepaspectratio, width=\linewidth]
    {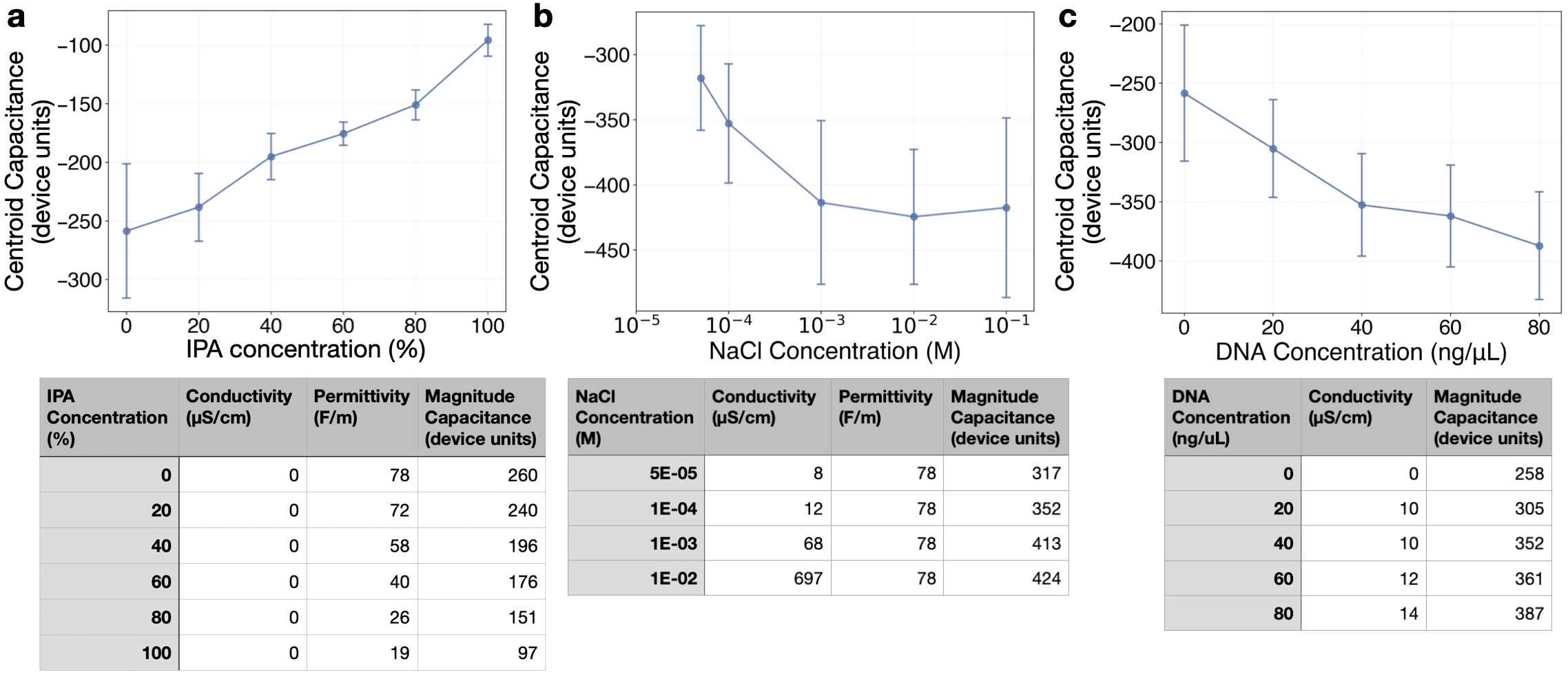}}
\caption{{\bf Effect of liquid concentration on capacitance readings.} {\bf (a)} Isopropyl alcohol (IPA) {\bf (b)} NaCl (salt water) {\bf (c)} DNA from calf thymus.}
\label{fig:theory1}
\end{figure}

\begin{figure}[t]
\centering
{\includegraphics[keepaspectratio, width=.6\linewidth]
    {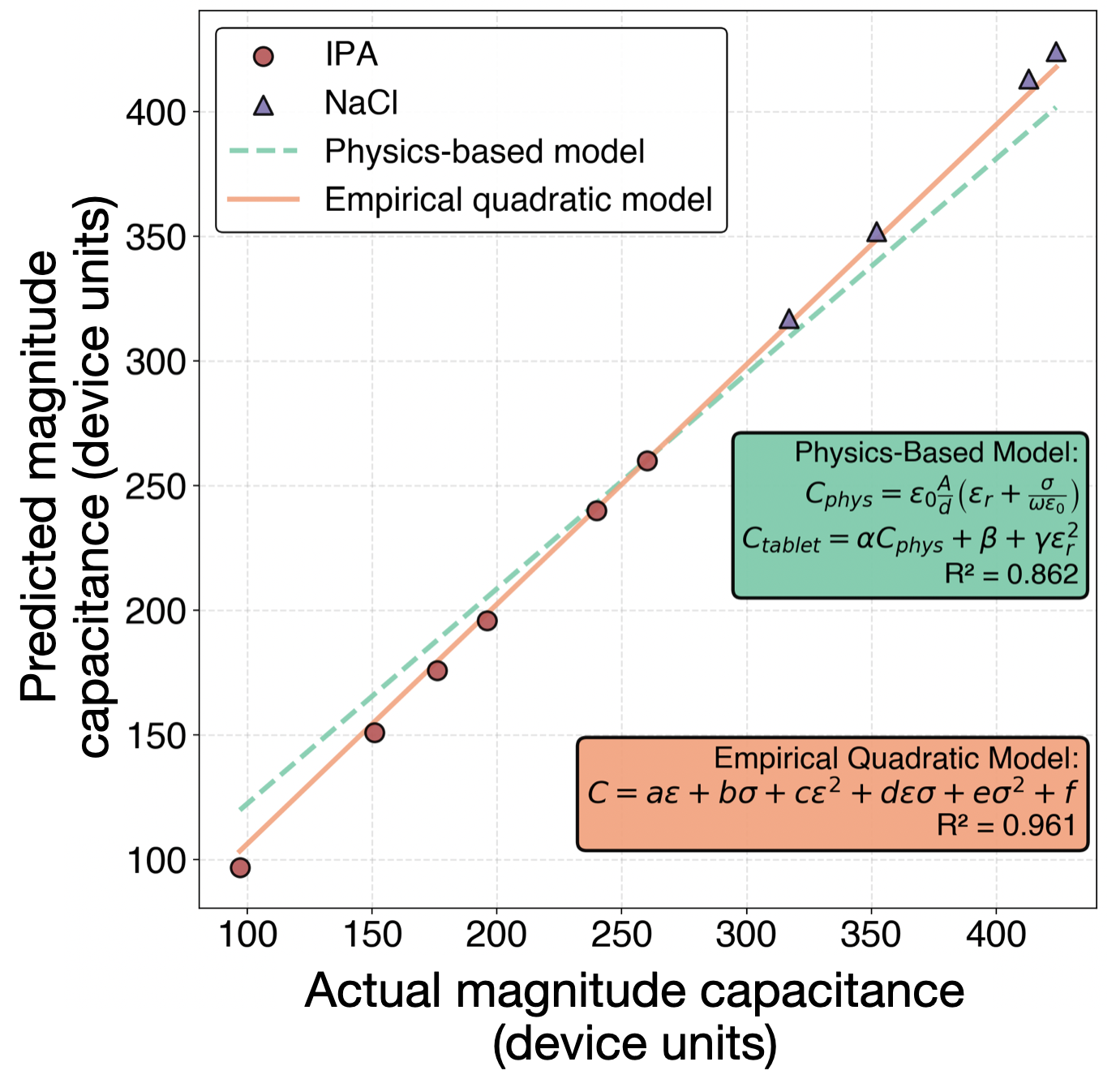}}
\vspace{-1em}
\caption{{\bf Modeling capacitance readings on tablet touchscreen.}}
\label{fig:theory2}
\end{figure}

With the filter disabled, we now quantify how liquid properties of conductivity and permittivity contributed to the measured capacitance values. 

\noindent {\bf Characterization of Liquid Concentration.} We perform an experimental analysis by placing different concentrations of salt water (NaCl) and isopropyl alcohol (IPA) concentrations of deionized water on the touchscreen and measured the device's capacitance readings. Specifically, we used alcohol concentrations of 0, 20, 40, 60, 80, 100\% and salt water concentrations of $5 \times 10^{-5}, 10^{-4}, 10^{-3}, 10^{-2}$~M. The reason why we selected these two liquids is because increases in alcohol concentration decreases the liquid's permittivity while its conductivity is largely constant, while conversely increases in salt water concentration will increase its conductivity while its permittivity is relatively constant. This allows us to decouple the effects of conductivity and permittivity on the measured capacitance and separately analyze their influence on the touchscreen's response. 

We show in Fig.~\ref{fig:theory1}a,b the conductivity of each sample as measured using an off-the-shelf electrical conductivity meter~\cite{ec_meter}, as well as permittivity values obtained from {related literature~\cite{park2006interfacial}}. We note that we do not obtain the permittivity values experimentally, as this typically requires a microwave network analyzer~\cite{vna} with a dielectric probe~\cite{vna_probe} costing over a hundred thousand dollars such as used in prior liquid sensing works~\cite{dhekne2018liquid}, and is prohibitively expensive to acquire. We can observe in the table that the capacitance readings from the NaCl and IPA dataset also span non-overlapping ranges.

We show in Fig.~\ref{fig:theory1}a,b the measured capacitance reading as the IPA and NaCl concentrations increases. For each liquid we use drops of volume 500~\uL~and perform ten replicates. We can observe a general trend where the centroid value in device units increases with IPA concentrations, while it decreases and plateaus with NaCl concentrations. To further explore how liquid composition affects the capacitance response we examined aqueous solutions of DNA sodium salt from calf thymus~\cite{dna} at different concentrations of 0, 20, 40, 60, 80~ng/\uL, and show that the centroid capacitance values in device units decreases (Fig.~\ref{fig:theory1}c). 

We note here that while the mean values do show a trend for these three liquids, the error bars overlap, suggesting that using a single metric such as the centroid value is not sufficient for classifying between different concentrations, and leveraging additional spatial information is needed.

\noindent {\bf Physics-Informed Model.} We aim to fit a model $C_{tablet}(\sigma,\epsilon_r)$ that explains these trends, specifically focusing on the data points for the concentrations of IPA and NaCl. To do this, we first leverage a physics-informed model based on the idealized parallel-plate capacitor model Eq.~\ref{ref:eq1}. However, this idealized model alone is insufficient for two reasons:
\squishenum
\item  The touchscreen reports capacitance in arbitrary digital units, without a clear mapping to physical units of farads. 
\item In reality, there are non-ideal geometric effects~\cite{sloggett1986fringing,chew1980effects} and fringing field non-linearities (seen with the container in Fig.~\ref{fig:layers}).
% that need to be modeled using a term $\gamma \epsilon_r ^2$.
\squishenumend

To model these effects we assume that the mapping for touchscreen units is a linear mapping and assume a linear scaling factor $\alpha$ and offset $\beta$. While for the fringing effects we assume it is dominated by a quadratic term $\gamma \epsilon_r ^2$. Ultimately we model the capacitance as:

\begin{equation}
\begin{split}
C_{\text{screen}}(\sigma,\epsilon_r) &= \alpha C_{\text{physical}}(\sigma,\epsilon_r) + \beta + \gamma \epsilon_r^2 \\
&= \alpha \left( \epsilon_0 \frac{A}{d} \left( \epsilon_r + \frac{\sigma}{\omega \epsilon_0} \right) \right) + \beta + \gamma \epsilon_r^2
\end{split}
\end{equation}

% \item It does not account for capacitance loss due to the glass protective layer between the electrodes and the liquid, as well as any air gaps. In reality, capacitance is modeled in a series such as $\frac{1}{C_{total}}=\frac{1}{C_{glass}}+\frac{1}{C_{airgap}}+\frac{1}{C_{liquid}}$

% We show in Fig.~\ref{fig:theory2}) 
% Ultimately our physics-based model uses the model:
% \begin{equation}
% xxx
% \end{equation}

% \red{mention assumptions of A and d}

When fitted to the alcohol and NaCl water dataset, we find that Pearson's correlation coefficient is $R^2=0.862$ (Fig.~\ref{fig:theory2}). The fitted values are $A = 0.7081 mm^2$, $\alpha = 6.1360e+12$, $\beta = 101.8077$, $\gamma = 0.035408$.

We consider an alternative modeling approach which is entirely empirical and leverages a weighted combination of the conductivity and permittivity terms. We experimentally evaluate linear, quadratic, log-transformed, and power law based methods of modeling and find that a quadratic based modeling formulation produces the best fit

\begin{equation}
C_{\text{screen}}(\sigma,\epsilon_r) = a\epsilon + b\sigma+ c \epsilon^2 + d \epsilon \sigma + e \sigma^2 + f
\end{equation}

The result in Fig.~\ref{fig:theory2} shows that the empirical quadratic based model is able to fit the data with a Pearson's correlation of $R^2=0.974$ with fitted terms $a=4.147256$, $b=-0.318624$, $c=-0.007843$, $d=0.030901$, $e=-0.002721$, $f=12.1685$. This suggests that the empirical relationship between conductivity, permittivity, and measured device capacitance is non-linear, but that a 2nd-order polynomial formulation can characterize this relationship.

% \section{Characterizing touchscreen behavior}
\section{{\sysname} design}

\begin{figure}[t]
\centering
{\includegraphics[keepaspectratio, width=\linewidth]
    {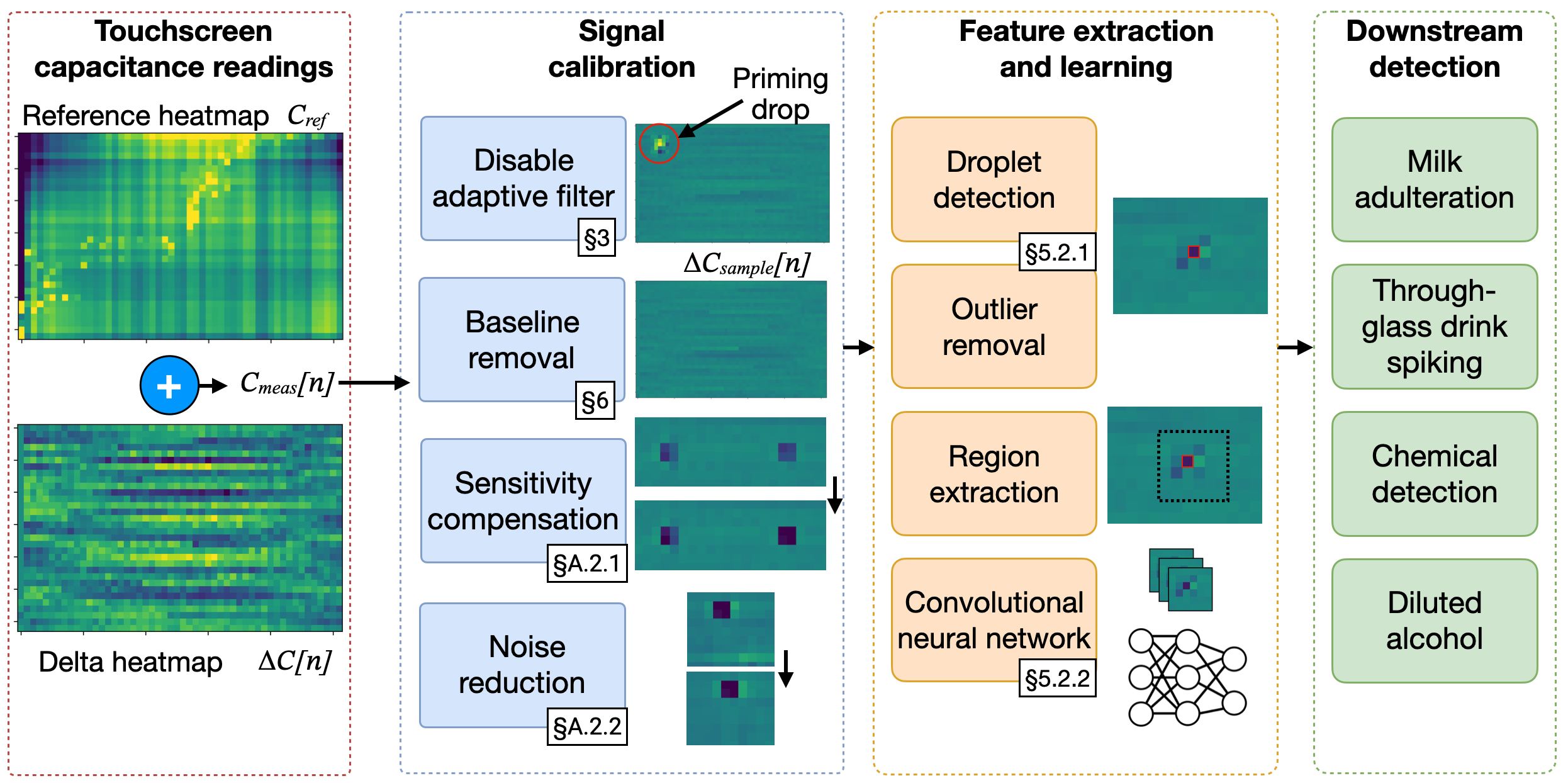}}
  \caption{{\bf {\sysname} system overview.}}
  % \vspace{-1em}
  \label{fig:pipeline}
\end{figure}

Now that we are able to perform liquid sensing, in this section we:
\squishenum
\item Perform characterization experiments to quantify the limits and behavior of the touchscreen sensor (Sec.~\ref{sec:analysis}).
% \item Present a physics-based and empirical model to model the device's response (Sec.~\ref{sec:theory}).
\item Present a learning-based approach to capture non-linear and spatial field effects (Sec.~\ref{sec:ml}).
\squishenumend

\subsection{Empirical characterization and analysis}
\label{sec:analysis}

\subsubsection{Effect of sample volume} 

\begin{figure}[h]
\centering
{\includegraphics[keepaspectratio, width=\linewidth]
    {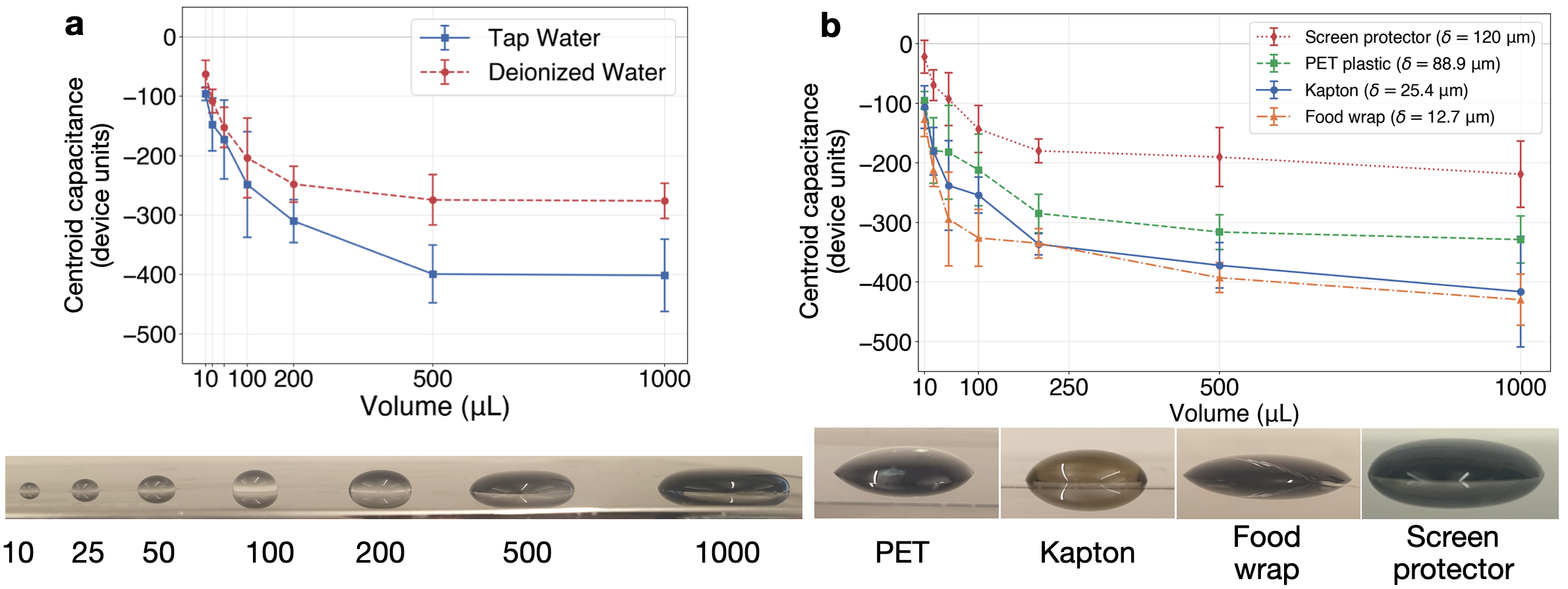}}
  \vspace{-2em}
  \caption{{\bf Effect of sample volume.} Measured across different {\bf (a)} water types {\bf (b)} thin-film insulating layers.}
  % {\bf (b)} sample temperature, and {\bf (c)} thin-film insulating layer. Note that these are the raw device values without compensation for spatial and temporal variation.}
  \label{fig:volume}
\end{figure}

We examined the effect of liquid volume on the measured centroid capacitance. We perform this experiment at volumes of 10, 25, 50, 100, 200, 500, and 1000~\uL~for tap water and deionized water~\cite{dei}. We perform 10 replicate measurements. Our results in Fig.~\ref{fig:volume}a show that as volume increases the magnitude of the capacitance increases, and stabilizes at 500~\uL~or higher. We select 500~\uL~ as the sample volume to use for our main evaluation to increase the separability between liquid classes.

% This trend arises because small droplets produce highly localized electric fields concentrated near the center of the drop, a phenomenon known as field focusing, which enhances signal magnitude but also increases variability when the drop covers only one or two electrode intersections. As the volume increases, the contact area broadens, averaging over more electrodes and reducing sensitivity to small placement errors or surface roughness. 

% xxx \red{should note that the weighted centroid capacitance is linearly decreasing..., say we decided to use 500 uL for our experiments...}

\subsubsection{Effect of Interfacial Layers} Here we consider the effect of two types of interfacial layers between the touchscreen glass and the liquid: thin-films and containers.

\noindent {\bf Thin-film Insulating Layer.} We examine the effect of adding a thin-film layer on top of the tablet touchscreen on the measured capacitance (Fig.~\ref{fig:volume}b). In principle having the least amount of separation between the liquid sample between the sample and the electrodes will yield the highest capacitance changes. However, one may wish to add an insulating layer when testing viscous liquids such as oil or milk, or biomedical liquid samples such as DNA. 

\begin{figure}
\centering
{\includegraphics[keepaspectratio, width=\linewidth]
    {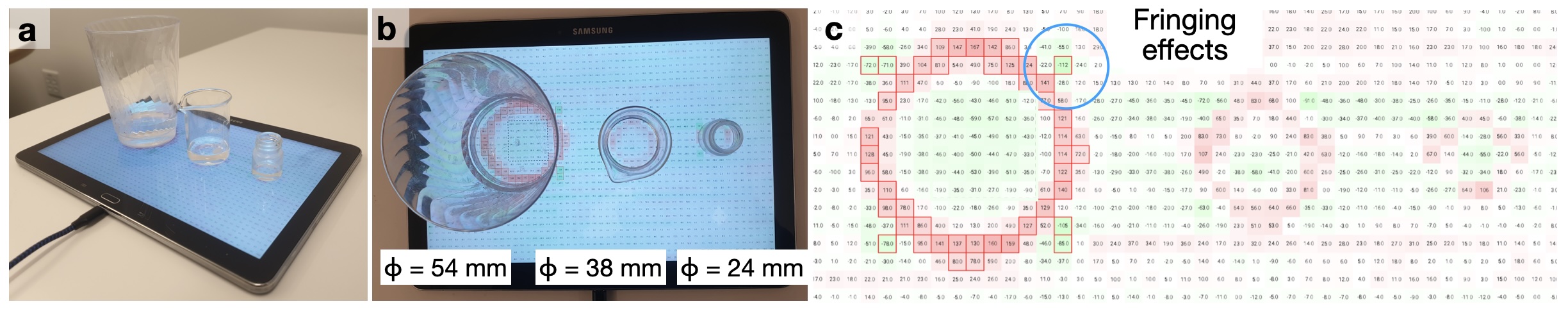}}
  \caption{{\bf Effect of different containers for through-container sensing.}}
  \vspace{-1em}
  \label{fig:layers}
\end{figure}

We used three films: food wrap ($\delta=12.7 \mu m$)~\cite{food_wrap}, Kapton tape ($\delta=25.4 \mu m$)~\cite{kapton}, and PET plastic ($\delta=88.9 \mu m$)~\cite{pet} (similar to material used in overhead transparency films~\cite{overhead}), where $\delta$ is the thickness of the film. We show the centroid capacitance for different volume levels of water. The results in Fig.~\ref{fig:layers}a show the capacitance change for thinner films is larger than with thicker films, with differences showing starting at 50~\uL~and becoming more pronounced at larger volumes up to 1000~\uL. 

The reason for this is that when an additional dielectric layer is added the total capacitance can be modeled as a series of capacitors representing the glass screen, the thin film, and the liquid sample:

\begin{equation}
\frac{1}{C_{\text{total}}} = \frac{1}{C_{\text{glass}}} + \frac{1}{C_{\text{film}}} + \frac{1}{C_{\text{liquid}}} 
\end{equation}

Because capacitances in series combine reciprocally, the presence of the thin film reduces the overall effective capacitance and attenuates the measured signal. 

\subsubsection{Through-Container Sensing.} Given that the electric field is able to pass through thin films with up to $88.9 \mu m$ as in the case of PET plastic, we next examine the effects of through-container sensing of liquids in containers (Fig.~\ref{fig:layers}). To do this, we select three different containers: a plastic drinking cup ($\phi=54~mm$), a beaker ($\phi=38~mm$), and a vial ($\phi=24~mm$), and filled them with water until they reached the 2~cm mark and placed them on the touchscreen. The measured heatmaps reveal four points:
\squishlist
\item The largest capacitance changes occur along the container's rim. This occurs because the base is slightly concave, causing only the perimeter to make firm contact with the screen while the center remains slightly elevated.
\item The perimeter exhibits negative capacitance (positive in device units), resembling the response from a finger touch, indicating that conductive coupling dominates here rather than permittivity/dielectric effects as seen with thin films.
\item In contrast, at the center of the container, capacitance increases (negative in device units), which is similar to the response observed with thin films dominated by the permittivity effect. 
\item There are fringing effects at the four corners around the cup even at electrodes where the cup makes no contact with the screen. This occurs because the electric field lines are not perfectly confined between adjacent electrodes and extend outward into the surrounding region. When a large container perturbs the field, these fringing lines couple capacitively to the container and the liquid inside it.
\squishend

% We next examine what happens the feasibility of whether the tablet is able to perform sensing through-container. We show in Fig.~\ref{fig:layers}b,c,d the capacitance heatmap generated when placing a plastic cup~\cite{plastic_cup} on it with 50~mL of tap water. 

% note that the values are weaker here
% note the fringing effects around the edges
% note the concavity of the container

% -> note that interestingly enough, it is acting like a conductor like a finger by default, why is that?
% but the sids has some fringing effects

% \Delta C = C_\text{mutual, new} - C_\text{mutual, baseline}
% C_\text{total} = C_\text{mutual} + C_\text{air} + C_\text{earth} + C_\text{parasitic}

% Through-container:
% We note multiple observations:
% (1) Liquid
% Having established how the touchscreen responds to physical liquid properties, we next design {\sysname} to leverage these effects for sensing.

\subsection{Feature extraction and learning}
\label{sec:ml}

Based on the findings from our empirical analysis and modeling, we design a feature extraction and learning-based pipeline for downstream liquid detection and classification applications.

\subsubsection{Automatic droplet identification}
\label{sec:detect}

\red{To automatically localize liquid droplets, we employ a $z$-score-based detection algorithm. For each frame, pixels with values below a negative threshold $\mu - z\cdot \sigma$ (with $z=2$) are identified as candidate droplet locations, reflecting the characteristic decrease in capacitance (in device units) caused by liquid contact. Connected components are then extracted using 8-connectivity, and small regions likely due to noise are removed based on a minimum size threshold.}

\red{Candidate regions are further filtered based on geometric constraints, retaining only compact regions with approximately elliptical shapes. To reduce unnecessary computation, droplet detection is only triggered when a significant frame-to-frame change is observed, indicating that a droplet has been newly deposited on the screen.}

\subsubsection{Capturing liquid spreading behavior}

Prior work on capacitive liquid sensing~\cite{horstmann2021capacitive} have
focused on reporting the capacitance value at the centroid of the liquid deposit location. \red{We note that using this alone is typically insufficient to distinguish between different liquids
and classifications, and error bars between different classes overlap. }

\red{Our approach instead extract capacitance measurements from an \dimens{8}{8} patch centered at the detected droplet location. This region is large enough to capture spatial spreading behavior while remaining computationally efficient. }\textit{This spatial information implicitly captures the effects of surface tension which affect how the liquid spreads across the surfaces and interacts with neighboring electrodes.} For example as shown in Fig.~\ref{fig:surface_tension}, when comparing water ($\gamma\approx72 ~mN/m$) and isopropyl alcohol ($\gamma\approx23~mN/m$), alcohol spreads and flattens more easily. So for the same volume, alcohol occupies a larger surface area and changes the capacitance in a larger neighborhood compared to water.

\red{We designed a compact CNN to comprehensively extract all these features effectively. The network consists of two convolutional layers with \dimens{3}{3} kernels followed by max pooling and dropout. This lightweight architecture effectively extracts discriminative spatial features from individual frames, enabling robust liquid and adulteration classification.
}
\section{Implementation}
\label{sec:impl}

\begin{figure}
\centering
{\includegraphics[keepaspectratio, width=\linewidth]
    {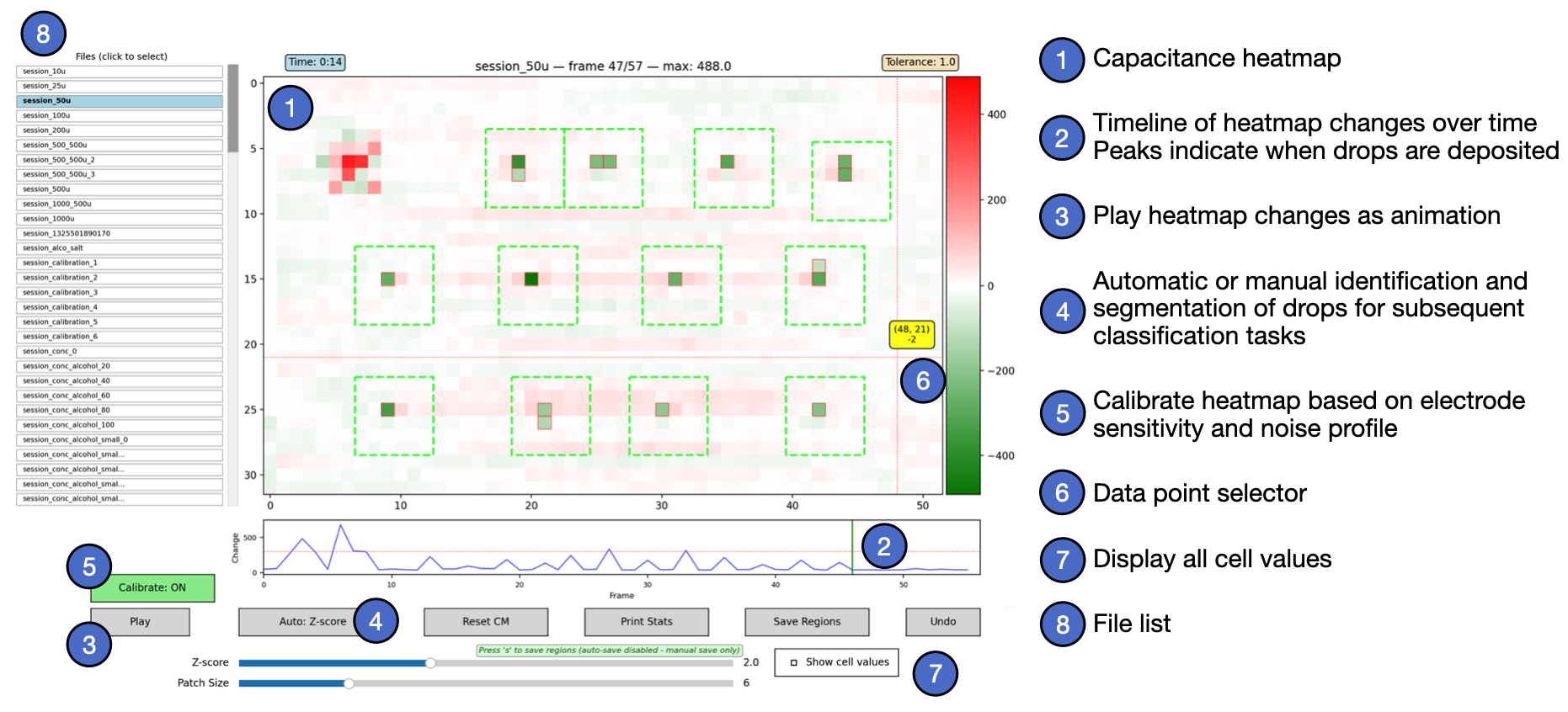}}
  \caption{{\bf Desktop app for visualizing capacitance measurements, tuning droplet segmentation parameters, and calibration algorithms to address sensitivity and noise.}}
  \vspace{-1em}
  \label{fig:sw}
\end{figure}

All of our experiments are performed on the Samsung Galaxy Note 10.1 (2014, SM-P600)~\cite{note}, which is rooted to provide super-user capabilities to access the underlying mutual capacitance values. The tablet uses an Atmel maXTouch mxt1664S display controller, which can be accessed using Atmel's open-source command-line utility, \textit{mxt-app}~\cite{atmel}. The Atmel utility reads the mutual capacitance measurement data via I2C communication, polling the register in a loop. It can read the matrices of the capacitance references and deltas which is the difference between the measured and reference capacitance values. The fastest it is able to read is at a period of 0.6~s between frames. The software produces a capacitance matrix of size 52 $\times$ 32 cells, given that the tablet screen has a physical size of 21.6 $\times$ 13.5~cm, this indicates an electrode pitch of 4.2~mm. It produces deltas and references of the heatmaps.
% and we describe in Sec.~\ref{sec:baseline} how those are processed.

For preprocessing, we measure the reference capacitance heatmap $C_\text{ref}$ once at the start of each measurement session. We then continuously measure the delta values $\Delta C[n]$ at each frame. The measured capacitance at frame $n$ is obtained as:

\begin{equation}
C_\text{meas}[n] = C_\text{ref} + \Delta C[n]
\end{equation}

As the dielectric system of the touchscreen also includes the protective glass screen on top and air gaps (which all have relative permittivity $\epsilon_r < 5$), which also affects the electric field, we remove this baseline by subtracting the first measured heatmap from all subsequent heatmaps, giving:

\begin{equation}
\Delta C_\text{sample}[n] = C_\text{meas}[n] - C_\text{meas}[0]
\end{equation}

\subsection{Software infrastructure}
We developed three pieces of software to assist with the design of our system:
\squishenum
\item {\bf Firmware Patch for Capacitance Access.} The current version of the utility provided by Atmel did not work on our version of the tablet as it had strict hardware validation checks for optional controller objects that would cause the app to terminate. We wrote a software patch to swallow these errors, log warnings and use default values instead. 
\item {\bf Visualization and Annotation Tool.} This is a Python app  (Fig.~\ref{fig:sw}) that allows viewing and segmenting the droplets from the heatmap file for downstream use in our machine learning algorithms. It provides features for automatic and manual labeling of droplet location, real-time adjustment of parameters, animation of the heatmap over time, and printing of different statistics about the selected droplet regions.
\item {\bf Real-Time Tablet Display App.} This is an Android app that calls the \textit{mxt-app} utility to write the capacitance values to disk, reads the recordings from disk and displays the values for each cell to the tablet screen. It implements computer vision thresholding algorithms to identify candidate droplet deposition locations, marks them, and runs them through a designated classifier to output in real-time the classification result on the screen. We find that a conservatively safe polling interval for a full cycle of write and read commands is 2.5~s.
\squishenumend

We will open-source this software infrastructure along with an end-to-end tutorial upon publication to enable others to build on top of these utilities.

% We note that there is a related method~\cite{nexus_hack} of accessing capacitance touch matrices on the LG Nexus 5~\cite{nexus} using a custom Linux kernel designed to access the Synaptics ClearPad 3350 touchscreen driver that exposes a 15 $\times$ 27 capacitive image at a rate of 20 frames per second and a similar pitch of 4.1~mm.

\subsection{Training methodology}

% mention that the train-test split methodology is similar to prior ubiquitous computing work using video frames~\cite{chan2022testing}

\red{We employ drop-wise five-fold cross-validation to prevent data leakage. Data partitioning between train and validation sets are conducted at the drop-level. Specifically, all time frames extracted from a single drop are assigned exclusively to either the training or validation set, never both, with an 80-20 train-test ratio across all samples. This ensures that frames from a single sample do not appear in both splits, and avoids optimistic bias.}

\red{To ensure sufficient data for training and validation, we extract 50 time frames per region. The classifier is trained and evaluated using 5-fold cross-validation, with training parameters set to 50 epochs, batch size 16, learning rate 0.001, and a ReduceLROnPlateau scheduler.}
\section{Evaluation}
\label{sec:eval}

\subsection{Adulteration detection}

\begin{figure}[h]
\centering
{\includegraphics[keepaspectratio, width=\linewidth]
    {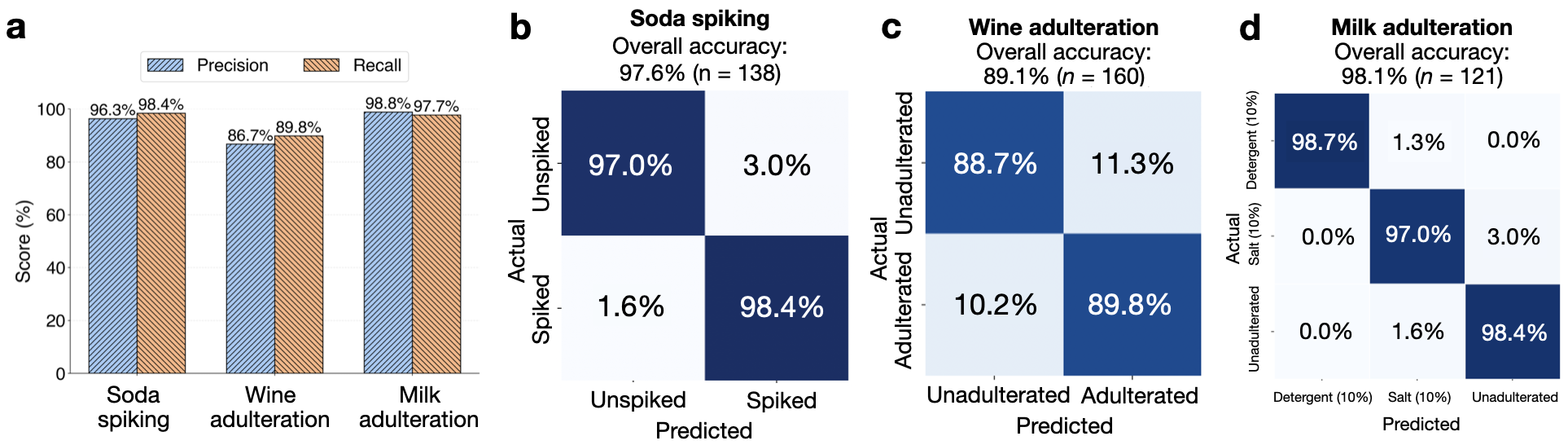}}
  \caption{\red{{\bf Liquid adulteration detection.} {\bf (a)} Precision and recall for detecting the adulterated liquid. The performance for both metrics exceed 85\% for the three use cases. {\bf (b--d)} confusion matrices for detecting adulterated liquids.}}
  \label{fig:binary}
\end{figure}

{Here we evaluate if our system is able to distinguish adulterated liquids and we outline the three scenarios below:}
\squishlist
\item {\bf Soda spiking.} Here we consider the scenario of a soft drink being spiked. To do this, we use as our base soda Coke, and spike it with \red{10\% volume concentration of industrial ethanol}, a form of alcohol unsuitable for human consumption. This scenario simulates intentional contamination with toxic substances. The dataset includes 77 unadulterated and \red{61} adulterated samples.
\item {\bf Wine adulteration.} Next, we investigate the adulteration of wine, which already contains alcohol. Here, the base liquid is red wine, to which industrial ethanol is added \red{at 20\% volume concentration}. Compared to soda spiking, this scenario is more challenging because the background alcohol content in wine reduces the relative change caused by adulteration. In total, 88 unadulterated and \red{72} adulterated samples are collected.
\item {\bf Milk adulteration.} We further explore milk adulteration, a common issue in dairy products where substances like salt and detergent are sometimes added to mimic the taste or appearance of pure milk. Salt can mask dilution, while detergent creates artificial froth but poses health risks. To simulate such cases, we mix 2~g of salt into 20~mL of milk and 2~mL of detergent into 20~mL of milk. The total numbers of unadulterated and adulterated samples are 44 and 74 respectively.
\squishend

Our results in Fig.~\ref{fig:binary}a show the precision and recall at detecting the adulterated class. We note that these metrics of interest better capture the expected performance compared to accuracy because they emphasize the model's ability to correctly identify adulterated samples even if it is at the cost of occasional false positives. In our application, the occasional false positive is acceptable as it may simply warrant re-testing, or in the worse case, discarding of the liquid sample. The converse is not acceptable as missing an adulterated product poses a greater risk than a false alarm. %Our results show our precision and recall is above 90\% for the three adulteration detection tasks.

We show the confusion matrix performance for the three tasks in Fig.~\ref{fig:binary}b--d, the results show that the overall accuracy of these classifiers are all above \red{89\%}. In the case of soda spiking the rate of false positives is less than \red{2\%}. For wine adulteration detection, we note this is a more challenging case as wine already contains alcohol so when it is contaminated with certain concentrations of alcohol, it is more challenging to detect the adulterated sample, however the overall accuracy is \red{89.1\%}. In the case of milk adulteration, the overall classification accuracy reaches 98.1\%. Notably, for detergent adulteration, which poses genuine health risks to humans, the false detection rate is 0.0\%, indicating that the system can reliably distinguish even subtle forms of contamination. \rednew{These results collectively demonstrate that capacitive touchscreen sensing can reliably distinguish adulterated from unadulterated liquids under controlled laboratory conditions with known adulterant types at relatively strong concentrations. We note that these scenarios represent a first feasibility demonstration and practical deployment would require evaluation across a broader range of adulterant types, concentrations, and operating conditions.}

% These results collectively demonstrate that our method provides a robust and reliable approach for detecting adulteration in various beverages and liquid foods.

\subsection{Concentration detection}
\label{sec:conc}

\begin{figure}[h]
\centering
{\includegraphics[keepaspectratio, width=\linewidth]
    {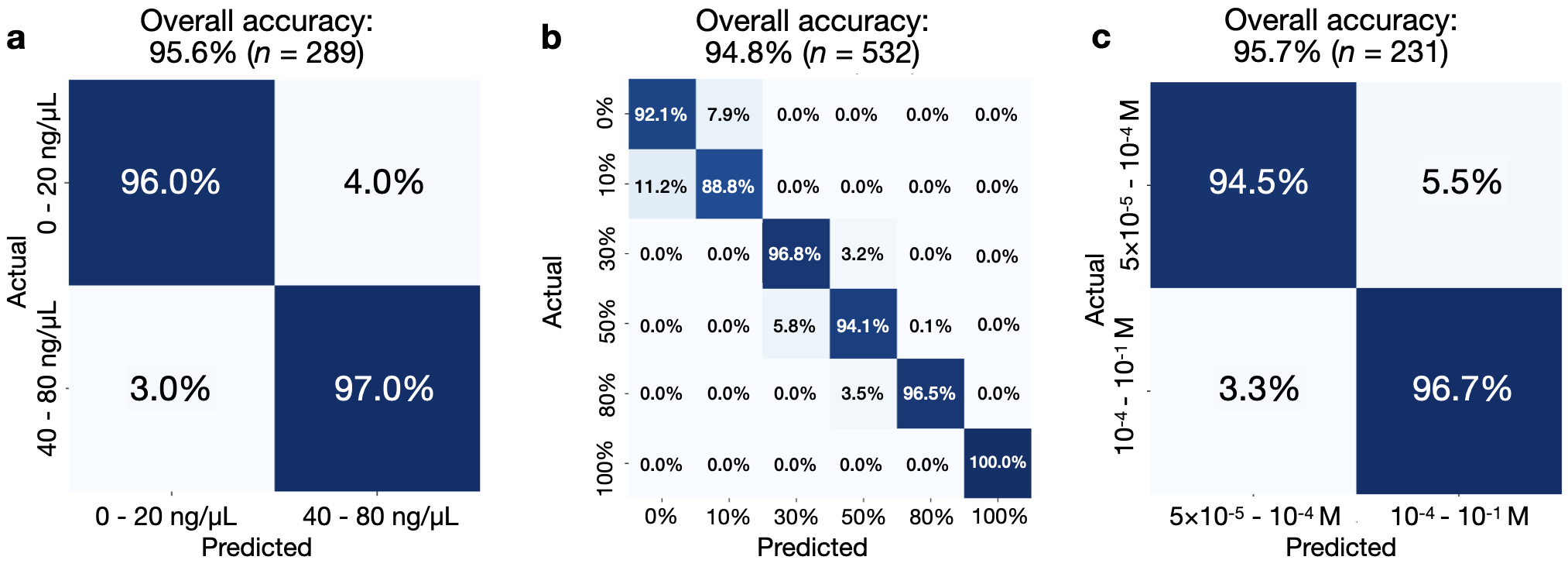}}
  \caption{{\bf Liquid concentration detection.} Confusion matrices for {\bf (a)} detecting trace concentrations of DNA (calf thymus) at a threshold of 20~ng/\uL, {\bf (b)} ethanol concentration classification at increments of 10--20\%.  {\bf (c)} NaCl salt water concentrations at a threshold of $10^{-4}$.}
  \label{fig:conc}
\end{figure}

Here, we evaluate the extent to which our system could be used to distinguish between different concentrations of liquids. We consider distinguishing between different concentrations of DNA and NaCl salt water using the same concentration breakdown as in Fig.~\ref{fig:theory1}. We also perform this for ethanol with slightly different concentrations, though still covering the full 0 to 100\% range. The goal of this particular analysis is to determine the granularity at which these concentrations can be detected. 

We find that in the case of DNA and salt water (Fig.~\ref{fig:conc}a,c) our system is primarily effective as a binary detector of trace concentrations of liquid. \red{We note that in these experiments the touchscreen measures the combined capacitance of both the analyte and the buffer solution. However, any changes in capacitance across dilutions are due to varying analyte concentration.} Specifically in the case of DNA it is able to consider concentration of 0--20~ng/\uL~as a low concentration and $>$ 40~ng/\uL~as a high concentration and distinguish between those, essentially functioning as a detector with a threshold of 20~ng/\uL, yielding an overall detection accuracy of 95.6\%. Similarly, in the case of salt water, the touchscreen is able to function a detector of the liquid with a threshold of $10^{-4}$~M with an overall detection accuracy of 95.7\%. In the case of ethanol, we find that our system is able to distinguish between six different concentrations of alcohol specifically 0, 10, 30, 50, 80, 100\% with an overall accuracy of 94.8\% (Fig.~\ref{fig:conc}b). 

\subsection{Through-container sensing}
% \begin{figure}
% \centering
% {\includegraphics[keepaspectratio, width=.32\linewidth]
%     {figs/c1.jpg}}
% {\includegraphics[keepaspectratio, width=.32\linewidth]
%     {figs/c2.jpg}}
% {\includegraphics[keepaspectratio, width=.32\linewidth]
%     {figs/c3.jpg}}
%   \caption{{\bf Through-container} }
%   \vspace{-1em}
%   \label{fig:through_container}
% \end{figure}

\begin{figure}[h]
\centering
{\includegraphics[keepaspectratio, width=0.8\linewidth]
    {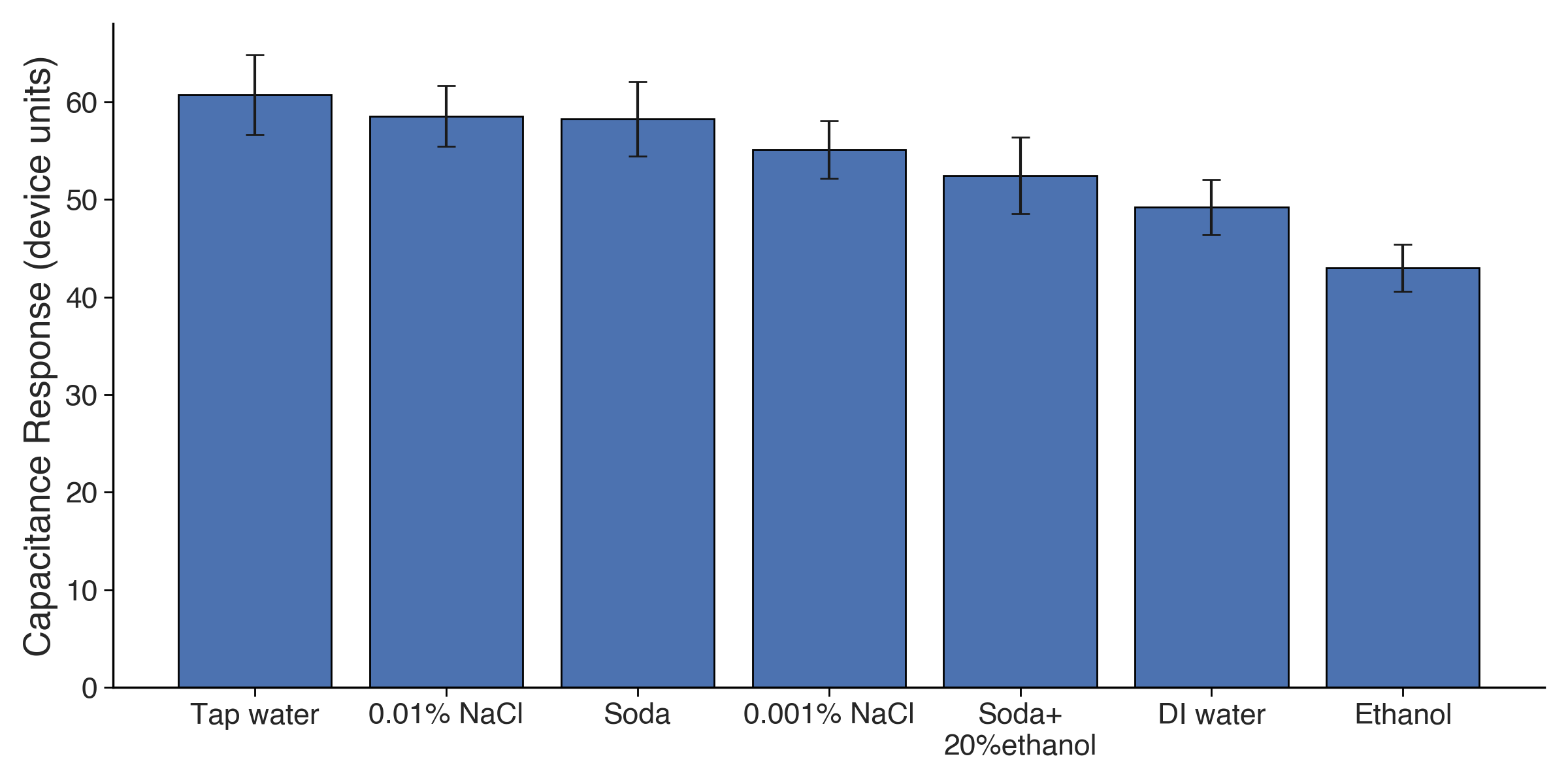}}
    \vspace{-1em}
  \caption{{\bf Through-container liquid adulteration detection.}}
  \label{fig:container_stats}
\end{figure}

\begin{figure}
\centering
{\includegraphics[keepaspectratio, width=\linewidth]
    {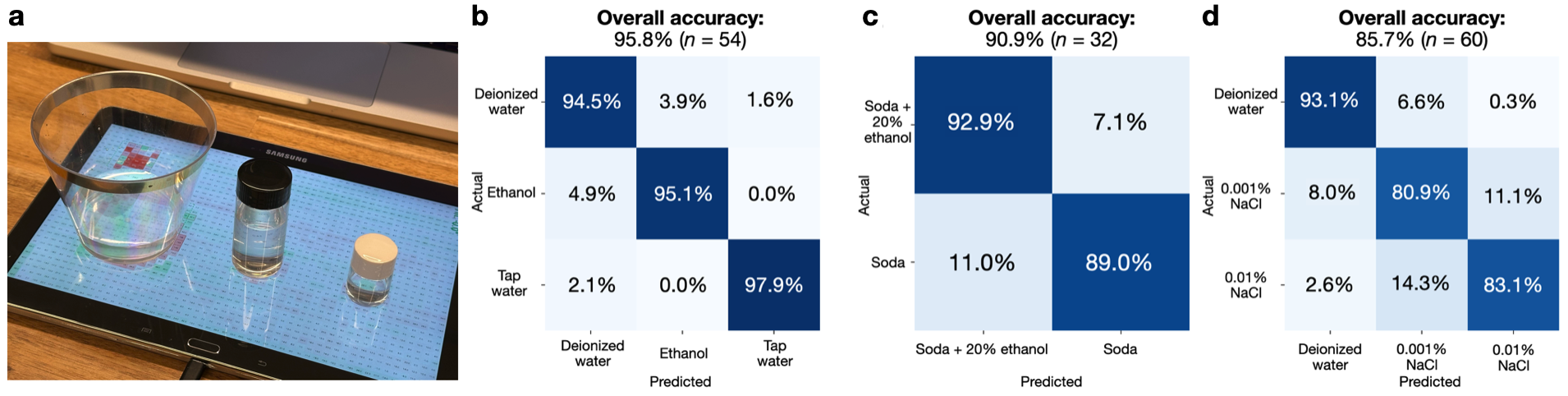}}
  \caption{{\bf Through-container liquid adulteration detection.} {\bf (a)} Experimental setup. {\bf (b--d)} Confusion matrices showing performance for different detection and classification tasks.}
  \label{fig:container_confusion}
\end{figure}

We further evaluate the system’s performance in through-container liquid concentration and adulteration detection, as directly dispensing liquids onto the touchscreen may not be practical in real-world use. In such cases, the base of the container introduces an additional dielectric layer that partially attenuates the capacitive response from the liquid in it.

% \begin{figure}
% \centering
% {\includegraphics[keepaspectratio, width=\linewidth]
%     {figs/container_bm.png}}
%   \caption{{\bf Through-container sensing benchmarks.} {\bf (a)} Experimental setup. {\bf (b)} Effect of container type and liquid volume.}
%   \label{fig:container_type}
% \end{figure}

As shown in Fig.~\ref{fig:layers}, placing a container on the screen produces a complex spatial response pattern, characterized by a strong positive capacitive change along the container’s rim, negative values at the center, and fringing effects near the corners. We observe that the rim region exhibits the most pronounced and consistent change across different liquids and container types. Therefore, we use the mean positive value along the rim as the primary feature for distinguishing between containers. We also used a 9~oz plastic cup as we find it generates the strongest capacitive response on the touchscreen due to its thin base.

We evaluated the through-container sensing capability to distinguish between different types of liquids. Three classification tasks were conducted:

\squishenum
\item \textbf {Basic liquid classification:}
Tap water, deionized water, and pure ethanol were tested to verify that the model can distinguish liquids with large differences in capacitance. The number of samples for tap water, deionized water, and pure ethanol were 18, 20, and 16, respectively.

\item \textbf{Soft drink adulteration detection:}
To evaluate the model’s ability to identify beverage adulteration, we used soda as the base liquid and adulterated it with 20\% industrial ethanol. Both unadulterated and adulterated soda classes contained 16 samples.

\item \textbf{NaCl concentration detection:}
To test sensitivity to liquid concentration, we compared deionized water, 0.001\% NaCl solution, and 0.01\% NaCl solution, each with 20 samples.
\squishenumend

Initially, we applied the same CNN model used in the drop-level liquid classification. However, due to the larger input dimension (approximately 13×13 pixels compared to 6×6 or 8×8 previously) and the more complex spatial pattern introduced by the container, the CNN struggled to learn discriminative features effectively, likely also affected by the limited dataset size. To address this, we adopted a feature-based random forest classifier using manually selected statistics from the region’s positive responses: the mean, median, and 75th percentile of positive values. We show the positive mean of different liquids tested in our through-container evaluation in Fig.~\ref{fig:container_stats}.

% As shown in Fig.~\ref{fig:container_stats}, the positive mean serves as a particularly effective metric for distinguishing between different liquids.

As shown in Fig.~\ref{fig:container_confusion}, through-container sensing achieved 95.8\% accuracy in the tap water–DI water–ethanol task, 90.9\% in soda adulteration detection, and 85.7\% in NaCl concentration classification. We note that these accuracies are slightly lower than when the liquid drop is directly on the screen, however this is expected as signal attenuation from the container's base can create more variation. However we note that the performance is still high exceeding 90\% on two of the tasks.

\subsection{Benchmarks}
\label{sec:benchmarks}

\begin{figure}
\centering
\includegraphics[keepaspectratio, width=.4\linewidth]
    {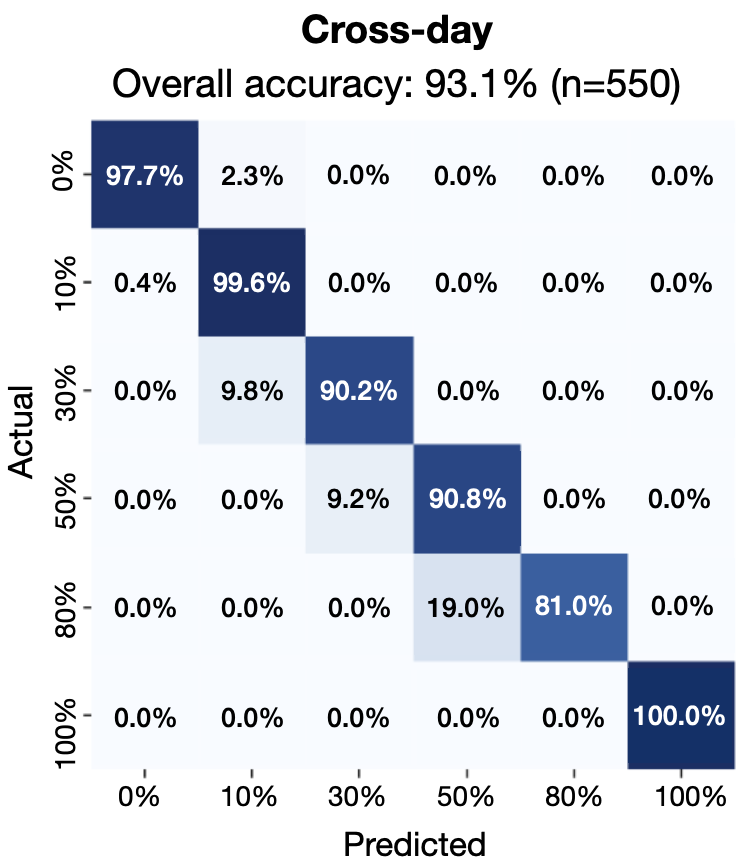}
  \caption{\red{{\bf Ethanol concentration classification evaluated across 74-day separation.}}}
  \vspace{-1em}
  \label{fig:cross-day}
\end{figure}

\noindent \red{{\bf Cross-day validation.} We conduct a cross-day evaluation using the multi-class ethanol concentration classification task described in Sec.~\ref{sec:conc}. In this evaluation, training and testing data are collected on different days separated by a 74-day interval, thereby capturing both session-to-session variation and longer-term environmental changes. All test samples originate from an unseen measurement day and unseen sessions. In both training and testing, droplet samples are placed at randomized locations across the screen, to capture variations in sensitivity across spatial regions. The dataset composition for training and testing is summarized in Table~\ref{tab:crossday_data}.}

\red{As shown in Fig.~\ref{fig:cross-day}, the cross-day evaluation achieves an overall classification accuracy of 93.1\%, comparable to the same-day training setting of 93.0\% shown in Sec.~\ref{sec:conc}. These results suggest that the system maintains stable performance over extended time intervals and is robust to session-level variability and day-level environmental noise.}

\begin{table}[t]
\centering
\begin{tabular}{lccc}
\toprule
{\bf Device} & {\bf Pitch size (mm)} & {\bf Controller} & \makecell[l]{{\bf Touch integration technology}} \\
\midrule
Samsung Galaxy Note 10.1 & 4.2 & Atmel mXT1664s & On-cell \\ \hline
LG Nexus 5 & 4.1 & Synaptics S3350B & In-cell \\ \hline
\makecell[l]{ATEVK-MXT2952TD-A\\ maXTouch Evaluation Kit} & 4.8 & Microchip ATMXT2952TD & N/A \\
\bottomrule
\end{tabular}
\caption{\red{Specifications of touchscreens used in our experiments.}}
\label{tab:screen_specs}
\end{table}

\begin{figure}
\centering
\includegraphics[keepaspectratio, width=\linewidth]
    {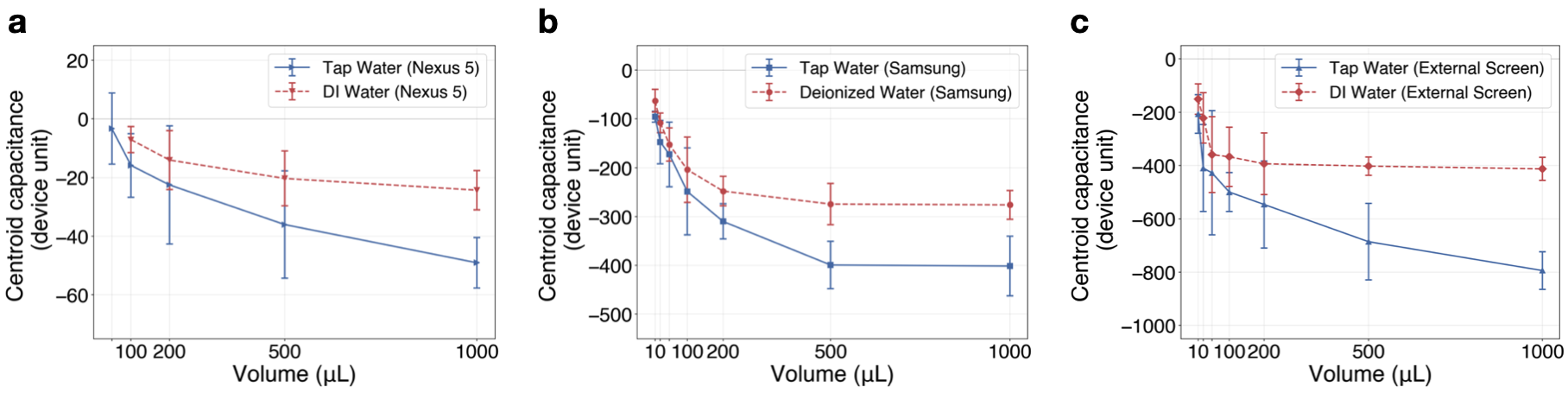}
\caption{\red{\textbf{Capacitance response across devices. } \textbf{(a)} LG Nexus 5. \textbf{(b)} Samsung Galaxy Note 10.1. \textbf{(c)} External screen (ATEVK-MXT2952TD-A maXTouch Evaluation Kit).}}
  \vspace{-1em}
  \label{fig:cross-device}
\end{figure}

\begin{figure}
\centering
\includegraphics[keepaspectratio, width=.6\linewidth]
    {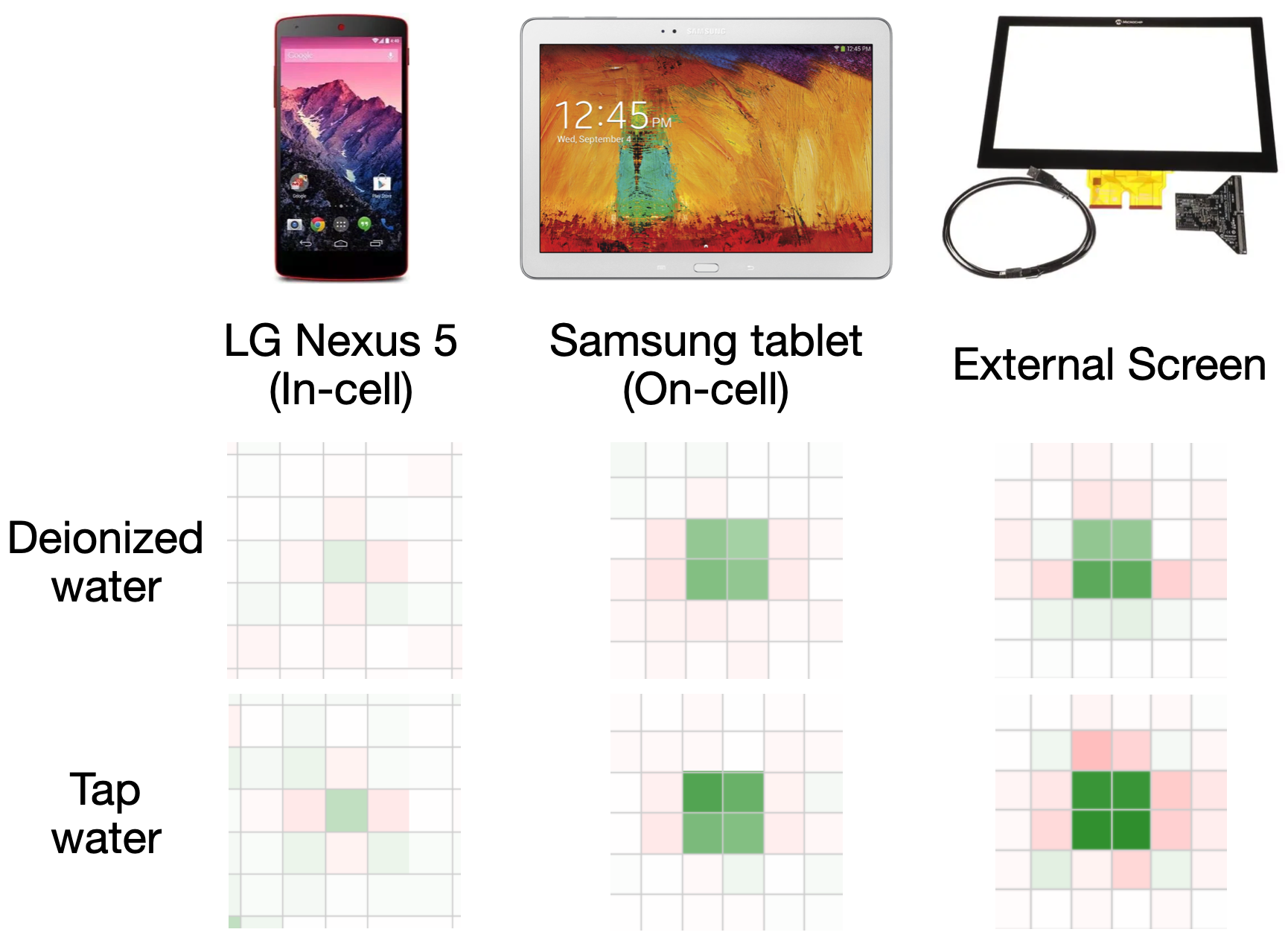}
\caption{\red{\textbf{Capacitance heatmap of 200~\uL~deionized water and tap water drops across three devices.}}}
  \vspace{-1em}
  \label{fig:cell}
\end{figure}

\noindent \red{{\bf Cross-device characterization.} To characterize our system's behavior across different devices, we tested three devices with different specifications: the Samsung Galaxy Note 10.1 tablet, the LG Nexus 5 smartphone, and the ATEVK-MXT2952TD-A maXTouch Evaluation Kit. While all three devices have similar pitch sizes (4--5mm), they use different touch controllers and different touch integration technology (Table~\ref{tab:screen_specs}).}

\red{We deposited liquid drops of various volumes on each device and recorded the raw capacitance data. We show in Fig.~\ref{fig:cross-device} the capacitance response of different devices for different volume levels. The figure shows that each device reports different absolute capacitance values, which is expected since touch controllers use arbitrary device units that are not standardized. However, deionized water and tap water, which differ significantly in conductivity, remain separable across all devices, and the trends are consistent. Due to the Nexus 5's lower sensitivity, it cannot reliably detect drops below 50 µL for tap water and 100 µL for deionized water (Fig.~\ref{fig:cross-device}a). This suggests that these different touchscreens are all able to perform liquid sensing albeit to different degrees.}

\red{We observe that the touch integration technology used by each device affects its sensitivity for liquid sensing. In-cell displays integrate touch sensing within the LCD layer, requiring the capacitive signal to pass through the liquid crystal material, which reduces sensitivity. On-cell displays place the touch sensor above the LCD but below the cover glass, providing better sensitivity. The external evaluation kit has no display stack, offering direct capacitive coupling and the highest sensitivity. 
The difference in sensitivity is also reflected in Fig.~\ref{fig:cell}: for identical 200 µL drops of deionized and tap water, the external screen shows the largest capacitance change, followed by the Samsung tablet (on-cell), and then the Nexus 5 (in-cell).}

\red{Finally, the standard priming drop technique works effectively on both the Samsung tablet and external screen. The Nexus 5's lower sensitivity requires a modified approach: spreading a small water drop with a finger to form a thin conductive film. While the execution differs, the underlying physics principle is the same: creating a conductive layer that mimics a persistent touch input.}

\begin{figure}
\centering
\includegraphics[keepaspectratio, width=.7\linewidth]
    {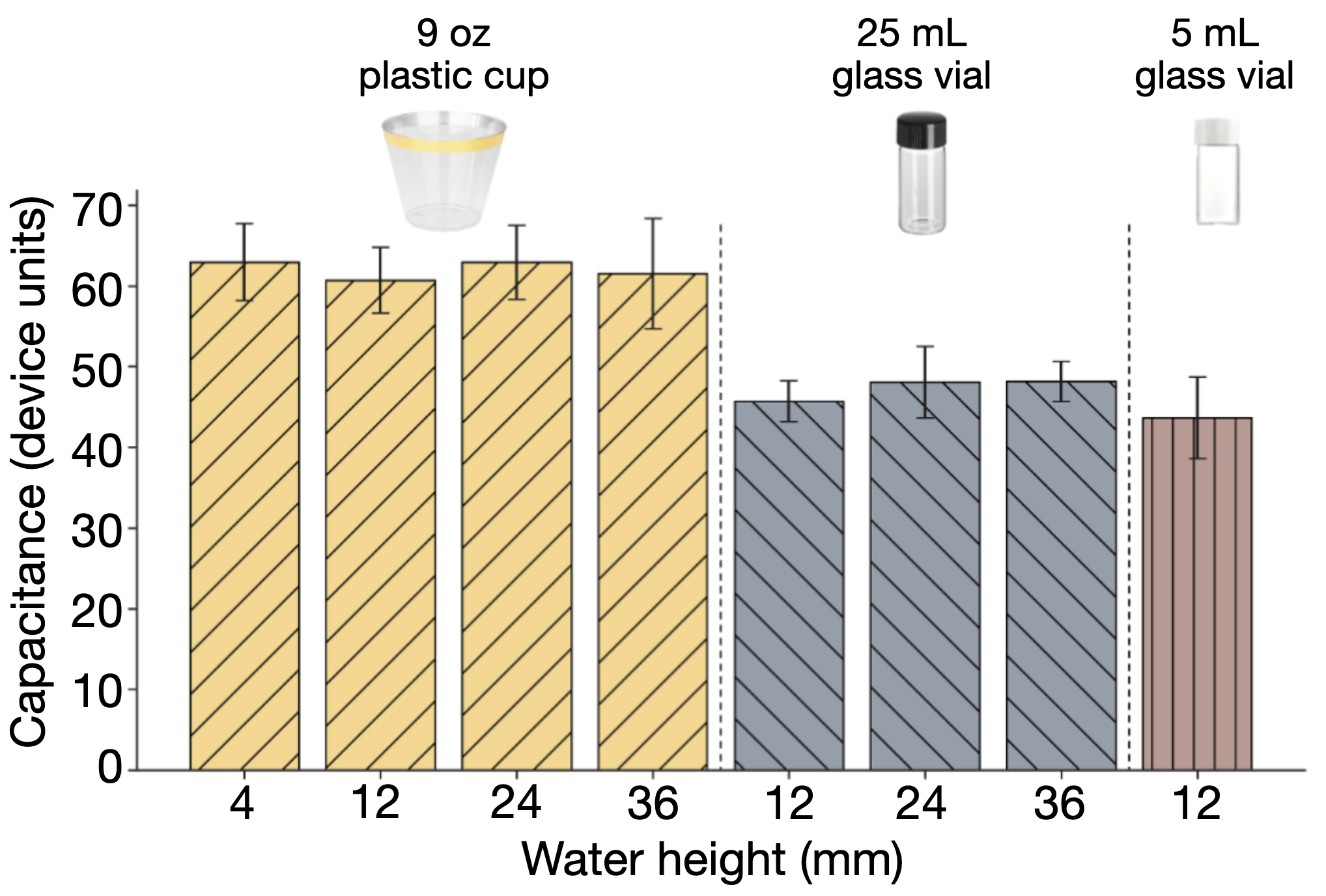}
  \caption{{\bf Through-container benchmarks}}
  \vspace{-1em}
  \label{fig:bm2}
\end{figure}

\noindent {\bf Effect of container type and liquid volume.} We compare three types of containers: 9~oz plastic cups~\cite{plastic_cup}, 25~mL glass vials, and 5~mL glass vials~\cite{vial_small}, chosen for their relatively flat and thin bottoms, which allow stronger capacitive coupling with the touchscreen. To investigate the effect of liquid volume, we test tap water at heights of 12, 24, and 36~mm in the 9~oz plastic cup and 25~mL vial. For the smaller 5~mL vial, only the 12~mm water level is feasible. Additionally, we test a small 5~mL water volume in the plastic cup to isolate the effect of liquid quantity from container type. 

The results in Fig.~\ref{fig:bm2} show that, for a given container, changing the water level or liquid amount does not significantly alter the positive mean response. This suggests that the dominant factor influencing the touchscreen’s capacitive response is likely the container's base thickness and material, rather than the volume of liquid inside. Since the 9~oz plastic cup generated the strongest capacitive response on the touchscreen, we selected it as the standard container for subsequent experiments. We used 23~mL of liquid for testing, although 5~mL produced a comparable response, 23~mL represents a more practical amount for everyday use.

\subsection{User study}

\begin{figure}
\centering
{\includegraphics[keepaspectratio, width=\linewidth]{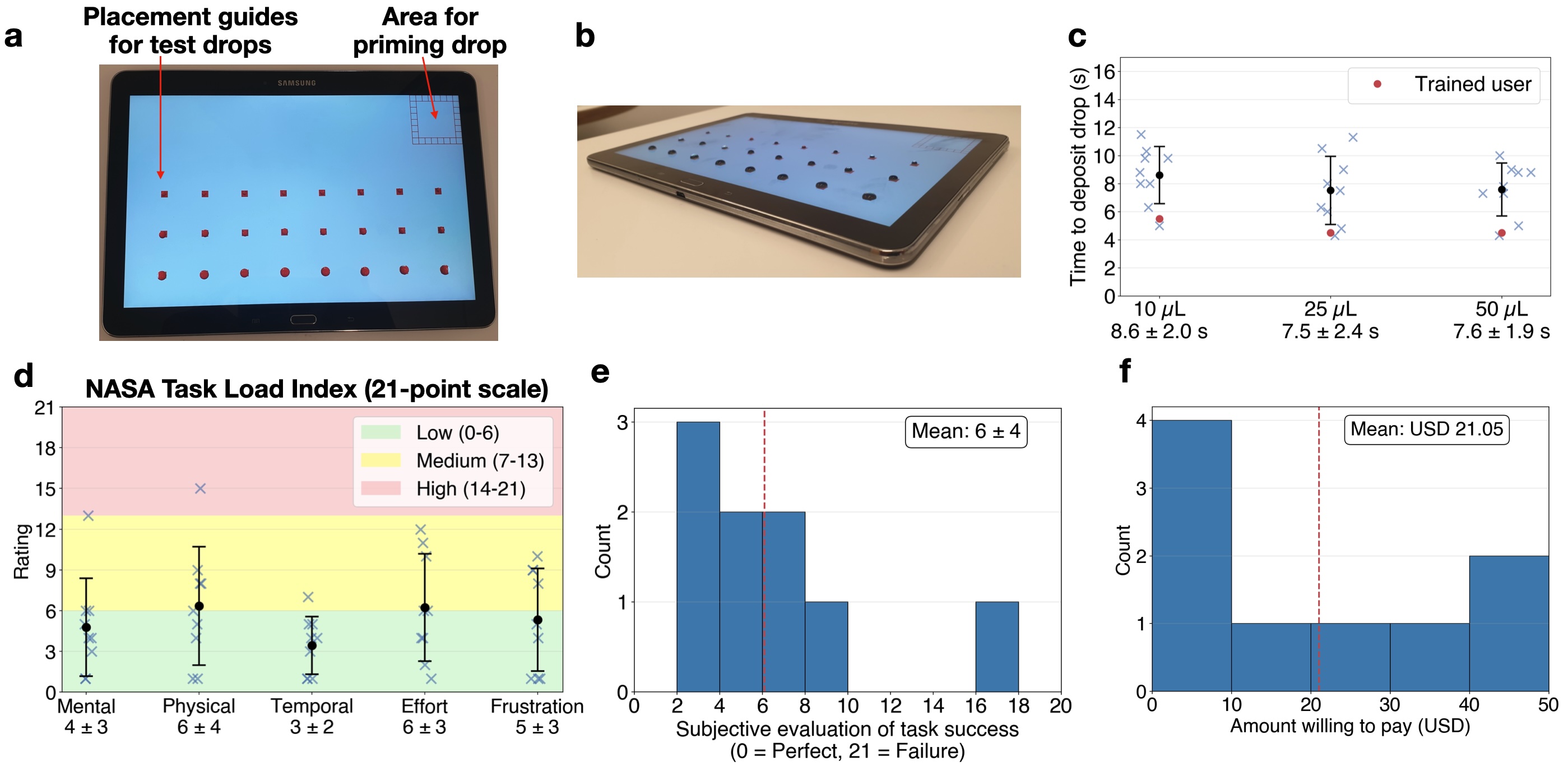}}
  \caption{{\bf User study evaluating different dimensions of the liquid sensing system ($n=9$).} {\bf (a, b)} Custom user interface with placement guides for the priming drop and test drops. {\bf (c)} Time to draw up and deposit test drops using a pipette for different volumes, comparing trained ($n=1$) and untrained ($n=9$) users. {\bf (d)} NASA Task Load Index assessing the system's workload across different dimensions on a 21-point scale. {\bf (e)} Users' subjective evaluation of task success. {\bf (f)} How much user's were willing to pay for the system.}
  % \vspace{-1em}
  \label{fig:usability}
\end{figure}

We conducted an user study determined exempt by our IRB (STUDY2025\_00000371) and recruited 9 participants by word of mouth from our institution. All subjects were PhDs students in computer science and engineering. We note that prior literature~\cite{nielsen1993mathematical} indicates testing on 5--8 participants is sufficient to uncover 85\% of usability issues in a system.

\noindent {\bf Study design.} The goal of this user study was to evaluate our system's ability to detect drops deposited by lay users, as well as the accuracy and time for user's to deposit drops on the screen. We also assess the system's workload and user's subjective evaluation of the system.
At the start of the study, we showed participants a demo of the real-time app illustrating that placing drops of liquids would change the capacitance heatmap. We then gave them a tutorial for how to use a pipette for the purposes of the usability study as none of them had used a pipette before. However, we note that lay users could always use eye drops which draw up a constant amount of fluid each time, and is effective at depositing the priming drop. We showed them the app interface (Fig.~\ref{fig:usability}a,b) which displayed an area to deposit the priming drop, and grid lines indicating individual electrodes, the center of which users had deposit the liquid sample of interest. The software marked a set of 24 cells split into 3 rows and 8 columns where the users had to deposit a drop. For the first row, users needed to deposit 8 drops of 10~\uL~across the 8 columns. This is repeated for the second and third rows where they had to deposit 25 and 50~\uL~samples.

\noindent {\bf Results.} Below we describe the results of our user study:

\noindent \textit{Drop detection rate.} We computed how many drops were detected by our software (Sec.~\ref{sec:detect}). In the 10~\uL~case, we find that for the 8 drops $\times$ 9 users = 72 drops, 67 of 72 drops (93\%) were detected by our algorithm, while for the 25 and 50~\uL~case, 71 of 72 (99\%) drops were detected by our algorithm. We note that our system provides visual feedback by highlighting the detected drop area, allowing it to prompt the user to deposit the drop within a specified time window and re-prompt if no drop is detected.

\noindent \textit{Accuracy of deposition location.} We computed the centroid of each drop and measured its deviation from the marked target cell. We find that across all but one detected drops, they were deposited precisely at the marked location. The single drop not correctly deposited only deviated from the correct location by one cell.
    
\noindent \textit{Time to deposit drops.} We measured the duration to deposit a row of 8 drops of 10, 25, and 50~\uL. We plot the average time to deposit a single drop in Fig.~\ref{fig:usability}c and show that for the 10~\uL~case users took on average 8.6 $\pm$ 2.0~s to deposit a drop which was approximately 1~s longer than the time to deposit 25 and 50~\uL~drops which took 7.5 $\pm$ 2.4~s and 7.6 $\pm$ 1.9~s on average respectively. This is likely because users found it more difficult to control the smaller 10~\uL~drops which required finer hand movements to release accurately onto the marked positions on the grid. We also show the time for a trained user to deposit the drop which ranged from 4.5 to 5.5~s, which is around 2--3~s faster than the average untrained user. Overall, we note that the time to deposit liquids onto the screen was relatively quick.

\noindent {\bf Post-task questionnaire.} Upon completing the task, we evaluated the subjective workload of using our system through the NASA Task Load Index (TLX)~\cite{hart2006nasa} where users were asked to assess workload across five scales which each had 21 gradations, they were asked:

\squishlist
\item How mentally demanding was the task?
\item How physically demanding was the task?
\item How hurried or rushed was the pace of the task?
\item How hard did you have to work to accomplish your level of performance?
\item How insecure, discouraged, irritated, stressed, and annoyed were you?
\squishend

The 21-point scale was categorized into low (0--6), medium (7--13), or high (14--21) effort categories. The results (Fig.~\ref{fig:usability}d) show that across all six scales, the average rating across the five task dimensions ranged from 3 to 6 which were within the low workload category. We note that some users rated some workload demands as medium and hard. We suspect this more likely reflects the use of a pipette to draw up and deposit liquid droplets, but we note that this procedure would be similar to using pipetting samples for use on lab-grade instrument. This scenario also reflects a more challenging use case of our system, as users can also use an eye dropper with a larger amount of liquid such as 500~\uL~which can be detected by our system. While our study focused on users placing the sample at a fairly precise location on the screen, our system is capable of sensing liquid through-containers as well.

Users were also asked three further questions, specifically:
\squishlist
\item How successful were you in accomplishing what you were asked to do? \\0 = Perfect, 21 = Failure
\item How satisfied are you with the system's overall ability to perform liquid sensing?\\Likert scale of 1 = Very dissatisfied, 5 = Very satisfied
\item How much would you be willing to pay for this system?
% \red{\item How would you imagine using this system?}
\squishend

Users' subjective evaluation of their success at the task as on average 6 $\pm$ 4, suggesting that on average users felt generally successful, but with minor issues (Fig.~\ref{fig:usability}d). The users' overall satisfaction with the system (Fig.~\ref{fig:usability}e) was on average 4.1 $\pm$ 0.7 which leans towards satisfied. Finally, users indicated they were on average willing to pay \$21.05 for the system, we note that this number also implicitly reflects spending preferences of our subject population.

%\red{
%Participants expressed strong interest in the application scenarios that our system could potentially enable. For instance, two participants noted that they could envision using such a system to support at-home diabetes monitoring, such as “placing a small drop of urine on the screen to help monitor diabetes indicators.” Another participant, after seeing the system’s ability to estimate NaCl concentration, suggested a cooking-related use case, remarking that they could imagine “putting a drop of soup on the screen to quickly check how salty it is.” In addition, one participant who avoids caffeine due to sensitivity commented that some beverages marketed as non-caffeinated still cause noticeable reactions, and speculated that “placing a cup on the tablet to check whether it contains trace amounts of caffeine” could be valuable. While these scenarios are exploratory and beyond the scope of the current evaluation, they highlight user interest in the broader potential of touchscreen-based liquid sensing.
%}
\section{Discussion}

\subsection{Limitations}
\noindent {\bf Detectable liquids.} As capacitance sensing depends on a material's conductivity and permittivity, liquids with low values for both, such as oils, are inherently difficult to detect. In particular oils have relative permittivity values $\epsilon_r$ that typically range from 1--2, which is substantially lower than water, which is 80~\cite{dhekne2018liquid}. In preliminary testing, we find that the screen is able to register drops of oil, however, the measured capacitance values are low. This is however a general limitation of capacitance-based sensing. Our system is therefore not intended as a universal liquid sensor. Instead, primarily targeted towards detection of liquids with contrasting dielectric values or concentration detection. 

\noindent {\bf Adulteration detection scope.} \rednew{The adulteration scenarios evaluated in this work use known adulterant types at fixed, relatively high concentrations. The present results should therefore be interpreted as a laboratory proof-of-concept demonstrating the feasibility of capacitive liquid sensing on commodity touchscreens, rather than a system validated for immediate real-world deployment in uncontrolled settings. Extending this work to lower adulterant concentrations, unknown adulterant types, and uncontrolled deployment settings remains an important direction for future work.}

\noindent \red{{\bf Container material, geometry, and thickness. } The containers that we selected for our through-container sensing experiments were made of plastic and had flat and thin bases to enable capacitive coupling between the touchscreen electrodes and the liquid sample. However, we note that containers made of other materials (e.g. glass, ceramic, metal) may exhibit different dielectric properties and require material-specific calibration. Furthermore, containers with irregular base geometries can produce distorted response patterns on the screen, while thicker bases may reduce the signal below the touchscreen’s detection threshold.}

\noindent \red{{\bf Broader device integration.} While our main system has been developed and evaluated primarily on one device, we characterized the capacitive response across three devices with different controllers and different touch integration technologies. Our cross-device experiments demonstrate that the underlying physics principles remain consistent across platforms, despite variations in absolute capacitance values and sensitivity. This suggests that while device-specific calibration is necessary, the sensing approach can be adapted to different touchscreen devices. }

We note that the field of ubiquitous computing and mobile systems has a long and rich history of recognizing the value of research leveraging specialized or restricted hardware capabilities available to only a fixed range of devices or requiring some level of software or hardware modifications, including near-infrared cameras~\cite{barry2022home}, LiDAR~\cite{chan2022testing}, high-frame-rate capacitive sensing displays~\cite{wang2019flextouch}, rooted devices with custom kernels to increase IMU sampling rates~\cite{laput2016viband}, thermal cameras~\cite{adhikary2024jouleseye}, or in-ear microphones on earbuds~\cite{cao2024earsteth,butkow2023heart,zhang2025lubdubdecoder}. In the same spirit, our work illustrates a plausible technical pathway for how tablet touchscreens could be used for liquid sensing.

% \noindent {\bf Experimental control factors.} We note that different physical factors affect the screen capacitance and need to be controlled to ensure a fair comparison between liquid samples. First, liquid capacitance varies based on temperature and humidity,  Second, sample volume affects the heatmap signature of the drop given that larger volumes spread over a larger area. Thus standardized environmental conditions and operating procedures would be needed for accurate and repeatable results, however this requirements are no different than for lab-grade devices.

\noindent \red{{\bf Sample drop volume.} Our experiments used a sample drop volume of 500 µL because it produces spatial patterns large enough relative to the touchscreen's electrode grid to make different liquids' capacitance features distinguishable. While the smallest detectable sample volume in our setup is approximately 10 µL, feature separability decreases as sample volume drops. Using touchscreens with denser electrode grids may improve separability at smaller volumes.} % R5

\noindent \red{{\bf Horizontal, flat surface as operating assumption.} Our system requires that the tablet be placed on a horizontal, flat surface, which is a reasonable assumption for our envisioned application scenarios in mostly controlled environments. This is because generalizing across tilting angles is an ill-posed physics problem. Specifically, even seemingly identical liquid drops on a uniform surface will exhibit different deformation and motion patterns when tilted due to contact-angle hysteresis and microscopic surface variations~\cite{Eral2013contact, bonn2009wetting}. As a result, it is challenging to practically compensate for tilt-induced behavior, which motivates our explicit operating assumption of a level surface.
}

\noindent \red{{\bf Cross-temperature compensation.} Temperature compensation must be liquid- and concentration-specific. Our system's response depends on three properties: conductivity, permittivity, and surface tension. In our benchmark experiment (Sec.~\ref{sec:app-bench}) we are able to use a linear model to compensate between tap and deionized water as both these liquids differ primarily in conductivity only, and conductivity is known to produce linear changes in temperature~\cite{Hayashi2004}. However, for liquids which differ in permittivity and surface tension even at ambient temperature, the situation is more complex as the capacitance readings will vary nonlinearly with temperature~\cite{uematsu1980static, persson2017dielectric,vazquez1995}.
}

\noindent \red{{\bf Screen protectors.}
As of 2024, more than 68\% of active smartphones globally use screen protectors~\cite{screenprotector}. We tested a commercially available screen protector (approximately 120 µm thick) and found that it attenuates the screen's capacitive response to liquids (Figure~\ref{fig:volume}(b)). The signal attenuation is significant enough that the priming drop does not function reliably. However, users can still perform measurements by maintaining continuous finger contact with the screen throughout the measurement process, which provides the necessary capacitive coupling. While this workaround enables functionality, it does reduce the convenience of our approach for users with thick screen protectors.}

\noindent \red{{\bf Priming.} 
The priming mechanism does not specifically require deposition tools such as a pipette or eyedropper. Users can simply wet the screen with a fingertip and remove the priming drop using a tissue via capillary action (Fig.~\ref{fig:finger}). Furthermore, priming is not strictly necessary, as users can maintain continuous finger contact on the screen to suppress the adaptive filter. While custom touchscreens and controllers are a viable alternative, our contribution lies in demonstrating liquid sensing on unmodified touchscreens.
}

\subsection{Future directions}
\noindent {\bf Leveraging onboard actuators to measure other liquid properties.} Future work could could focus on investigating if other onboard motors on the tablet such as the vibration motor and the backlight LCD/OLED could be used for measuring other liquid properties: 

\noindent \textit{Vibration motor.} In initial testing, we observed that when the tablet is balanced on top of a surface such as the motorized tilt platform in Fig.~\ref{fig:tilt}, the vibration motor generates sufficient force to agitate and gradually move small objects such as a 10~g weight~\cite{weights}, across the screen. This occurs when the material and geometric properties of the object resonates with the motor’s fundamental frequency. We note that the object does not move when the tablet is placed flat on the table, which dampens the effect of the vibration. 

We observed that depositing a drop of liquid on the screen and subsequently placing an object on it resulted in the liquid being smeared across the surface as the object moved. Further investigation is required to determine {\bf (1)} if these smearing signatures are correlated to other liquid properties such as viscosity and surface tension, {\bf (2)} how to design or select objects which are maximally resonant with the motor {\bf (3)} the effect of the object's initial position with respect to the motor on its trajectory. Given that this is an independent idea with its own set of technical challenges, we leave this as a future direction of investigation.

\noindent \textit{Display backlight.} Modern touchscreen displays typically provide backlight intensities of 500--1000 nits~\cite{nits} with up to 2000 nits in some cases. This illumination could potentially serve as an integrated optical excitation source for modulating the physical properties of liquid or solid samples placed on the screen. Such a capability may be particularly useful for characterizing photoconductive materials whose capacitance varies with light exposure.

An alternative formulation of such a system would be to have an external optical excitation source that can sweep over a larger optical wavelength range such as ultraviolet, visible-light, and near-infrared to impact multiple liquid samples on the screen simultaneously while using the screen as a reader. We leave this as a future direction for investigation.

\noindent {\bf Sensing of soil and solids.} Given the sensing capabilities of tablet touchscreens, characterizing the capacitive properties of different solids may be a useful further direction. In particular measuring soil moisture levels could be an interesting direction given that soil analyzers~\cite{soil1,soil2} often rely on capacitance-based sensing using parallel plate capacitors. We note that touchscreens are also sensitive enough to detect the presence of conductive materials such as copper (conductive tape), coins (nickel, copper, and zinc), and keys (brass, nickel) register, with these capabilities being used in prior works ~\cite{mayer2021super,villar2018project,xiao2016capcam,pedersen2011tangible,voelker2013pucs,voelker2015percs,chan2012capstones,kratz2011capwidgets,liang2014gaussstones,schmitz2017flexibles} on enabling tangible user interfaces and widgets that can interact with the touchscreen.

% \red{comment on how to deploy and train this}
\appendix

\renewcommand{\thesection}{\Alph{section}}
\renewcommand{\thesubsection}{\thesection.\arabic{subsection}}

\counterwithin{figure}{section}      % figures: A.1, A.2, ...
\renewcommand{\thefigure}{\thesection.\arabic{figure}}

\counterwithin{table}{section}       % optional: tables A.1, A.2, ...
\renewcommand{\thetable}{\thesection.\arabic{table}}

\counterwithin{equation}{section}
\renewcommand{\theequation}{\thesection.\arabic{equation}}

\section{Appendix}

\subsection{Primer on capacitive sensing on touchscreens}
\begin{figure}[h]
\centering
{\includegraphics[keepaspectratio, width=\linewidth]
    {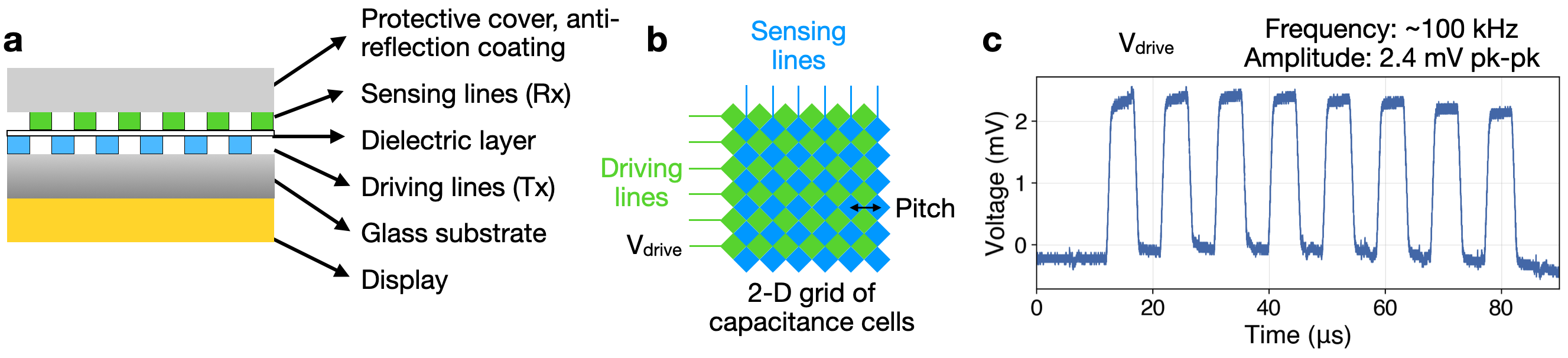}}
  \caption{{\bf Capacitive touchscreen architecture.} {\bf (a)} Layers of a capacitive touchscreen. {\bf (b)} Mesh of driving and sensing lines with a dielectric in between form a 2-D grid of capacitive electrodes or cells. The driving lines produce a voltage $V_{drive}$. {\bf (c)} $V_{drive}$ signal measured from an unmodified touchscreen tablet using an oscilloscope with probe in contact with the display. Figures are drawn for conceptual illustration and are not to scale.}
  \vspace{-1em}
  \label{fig:bg1}
\end{figure}

\noindent Modern multi-point touchscreens (Fig.~\ref{fig:bg1}a) are comprised of multiple layers starting from bottom to top:
\squishenum
\item {\bf Display:} Ranges from liquid crystal to light emitted diode based displays.
\item {\bf Insulating substrate:} Provides mechanical structural stability, generally glass.
\item {\bf Driving lines:} This is an electrode film, often polyethylene terephthalate (PET), with transparent electrode lines coated typically with indium tin oxide (ITO) which is known for its electrical conductivity, optical transparency, resistance to moisture, and ability to be deposited as a thin film. The touchscreen controller sends a driving AC signal along these.
\item {\bf Dielectric layer:} Usually PET or an optically clear adhesive (OHA) to mechanically bond and electrically isolate the top and bottom two ITO layers, prevents shorting between them. 
\item {\bf Sensing lines:} Similar to the driving line, but runs orthogonal or opposing to it to created a grid. It senses how much of the AC signal couples through the dielectric. The two electrode layers and the dielectric as a capacitor. 
\item {\bf Protective covering:} Usually glass with anti-reflective coating, which insulates the sensors from direct touch.
\squishenumend

Touchscreens operate on the idea of mutual capacitance which forms between the two electrode lines (Fig.~\ref{fig:bg1}b)~\cite{barrett2010projected,hessar2016enabling}. The screen structure is composed of a 2-D array of electrodes from the driving and sensing lines. On the driving line is an AC input drive signal $V_{drive}$, which creates a shared electric field between the layers, this is sensed by the other layer with the sensing electrodes. We show in Fig.~\ref{fig:bg1}c the driving signal of Samsung Galaxy Note 10.1 captured on an oscilloscope (T3DSO1302A, Teledyne LeCroy). The signal is captured by placing the oscilloscope probe directly in contact with the top of the touchscreen and connecting the ground to a metal pole. The driving frequency is approximately 100~kHz with a peak to peak distance is approximately 2.4~mV.

% \subsection{What the touchscreen measures}

\begin{figure}[h]
\centering
{\includegraphics[keepaspectratio, width=\linewidth]
    {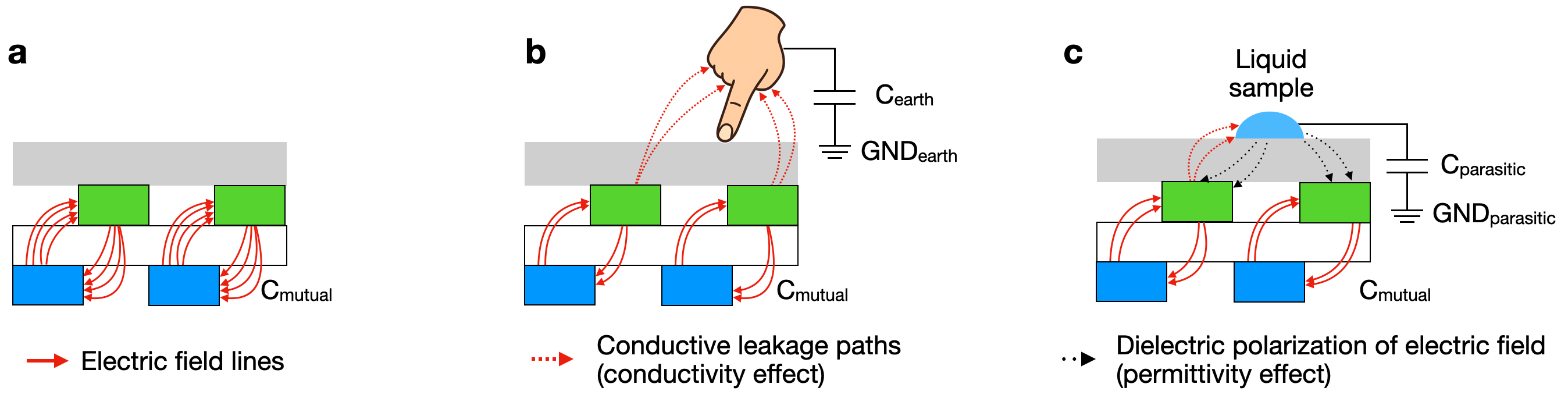}}
\caption{{\bf Capacitive sensing mechanisms under different interactions.} {\bf (a)} The two layers of driving and sensing electrodes form a mutual capacitance electric field $C_{mutual}$. {\bf (b)} The presence of a conductive grounded object like a finger creates a lower impedance path and electric charge is drawn towards it to $GND_{earth}$, this reduces the mutual capacitance field $C_{mutual}$ via the conductivity effect. {\bf (c)} When a liquid sample is placed on the surface it acts simultaneously as a conductor and a dielectric: as a conductor, the liquid draws away charge to $GND_{parasitic}$ as it is coupled to the air and other parasitics, while as a dielectric, it's relative permittivity $\epsilon_r$ increases its polarization creating a store of more energy and increases the effective capacitance between the electrodes and the liquid. The net change in capacitance is the sum effects of conductivity and permittivity.}
  \vspace{-1em}
  \label{fig:bg2}
\end{figure}

In a touchscreen without any object in contact with it, a mutual capacitive field forms between the driving and sensing electrodes (Fig.~\ref{fig:bg2}a). When a conductive object is near, it disturbs the shared electric field. In the case of grounded conductive objects like a human finger that is coupled to Earth ground, the electric field lines are pulled towards the finger and it \textit{reduces} the capacitance between two electrodes at an intersection (Fig.~\ref{fig:bg2}b). The touchscreen controller is able to read the change in capacitance and detect where it has occurred. The human body serves as an effective ground because it is capacitively coupled to Earth, forming a large conductive body with a low-impedance pathway to ground. For the electric field, this provides an easier route to discharge, as the finger is in closer proximity to the screen compared to some of the sensing electrodes which can be farther away and have higher impedance.

In contrast, when placing a liquid sample on the screen it alters the electric field coupling between the driving and sensing electrodes in two ways (Fig.~\ref{fig:bg2}c):

\squishenum
\item {\bf Conductivity effect.} Liquids containing free ions (e.g. salt water) act similar to the finger in that it can conduct small amounts of current. This provides a partial path to ground as it is parasitically coupled to the air and the device chassis itself. This lossy coupling causes part of the AC signal to leak through the liquid's conductivity. Under the parallel-plate capacitor approximation, the effective conductive contribution to capacitance can be expressed as:

\begin{equation}
C_{\sigma} = \frac{\sigma A}{\omega d}
\end{equation}

where $\sigma$ is electrical conductivity of the liquid, $A$ is the electrode area, $d$ is the separation distance or pitch between electrodes, and $\omega$ is the angular frequency of excitation $V_{drive}$. The remaining charge is redistributed to neighboring electrodes, \textit{effectively reducing the mutual capacitance readings~\cite{zimmerman1995applying}}.

\item {\bf Permittivity effect.} When liquid with a relative permittivity ($\epsilon_r$) higher than air contacts the screen, it acts as a dielectric that increases the degree of polarization in the electric field region. Specifically, what happens is that polar molecules align partially with the field, storing more electric energy and increasing the effective capacitance between electrodes even if the liquid is not strongly conductive according to:

\begin{equation}
C_{\epsilon_r} = \epsilon_r \epsilon_0 \frac{A}{d}
\end{equation}

where $\epsilon_0$ is vacuum permittivity. Non-conductive or poorly conductive liquids such as oils and alcohols influence the measured capacitance primarily through this effect \textit{creating an increase in the measured capacitance.}
\squishenumend

Overall, as a liquid sample can act as both a conductor and a dielectric and the measured capacitance on the display is both a function of both the liquid's conductivity and permittivity: 

\begin{equation}
C_\text{eff} = \epsilon_0 \frac{A}{d}\left(\epsilon_r + \frac{\sigma}{\omega \epsilon_0}\right)
\label{ref:eq1}
\end{equation}

% \red{mention here the concentration of salt and alcohol and do the capacitance readings...}
% We note however, that water drops on a screen a dominated by the permittivity effect. The reason for this is that the height of the liquid (governed by its surface tension) can be thicker than the touchscreen thickness (often around 1~mm), and so the electric field can be trapped

Building on these principles of capacitive sensing, the next section describes how they can be leveraged to enable liquid sensing on touchscreens.
% These physical interactions motivate our design for enabling liquid sensing on commodity touchscreens, which we describe next.

\subsection{Empirical characterization and analysis}

\subsubsection{Multi-scale spatial sensitivity analysis}
Here, we characterize the spatial sensitivity of the electrodes at the micro-scale, and at the macro-scale across the entire screen.

\noindent {\bf Micro-scale characterization.} Given that electric fields may not be uniform around an electrode, our goal here is to analyze where is the best position around a cell to deposit a drop. To do this, we deposited water samples of 10 and 25~\uL~at the center of an electrode, at the corner of the electrode which would be at the intersection of a drive and sense line, along a horizontal line and along a vertical line. We repeat this for 10 droplets for each of the volumes. 

\begin{figure}[t]
\centering
{\includegraphics[keepaspectratio, width=.9\linewidth]
    {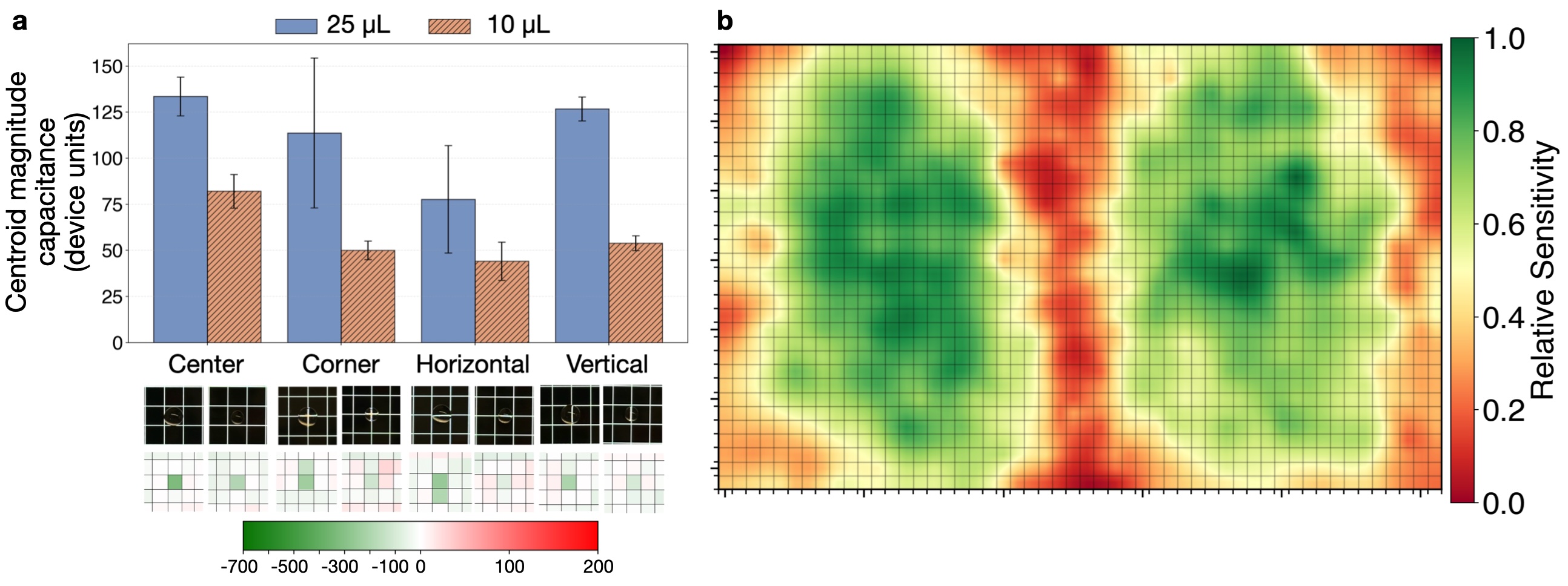}}
  % \caption{{\bf Spatial sensitivity analysis.} {\bf (a)} Detection accuracy of our droplet detection algorithm at different locations across ten replicates and two volumes. {\bf (b)} Magnitude capacitance of the 3 $\times$ 3 region around the centroid for different deposition locations.}
  \caption{{\bf Spatial sensitivity analysis.} {\bf (a)} Micro-scale analysis: magnitude capacitance of the 3 $\times$ 3 region around the centroid for different deposition locations. {\bf (b)} Macro-scale analysis: sensitivity across the tablet shows distinct regions of higher and lower response.}
  % {\bf (c)} Histogram of relative sensitivity values across all measured cells. A compensation map is computed as the inverse of the sensitivity map and applied to subsequent measurements to normalize spatial variation.
  \vspace{-1em}
  \label{fig:location}
\end{figure}

To characterize the change in capacitance instead of using the centroid which only measures the value at a single cell, which would make sense if the droplet was deposited only in the center. For drops in other locations which straddle the boundary between multiple cells we sum together the values across cells with a negative change in capacitance. We show in Fig.~\ref{fig:location}a the magnitude capacitance of the droplet. Both the 10 and 25~\uL~samples had comparatively higher capacitance values and lower variation at the center and vertical locations, and lower values at the corner and horizontal locations. The results suggest that placing droplets at the center of the electrode produce the most reliable results. This is the position that we adopt in all experiments. We note however that the use of larger volume samples occupy multiple cells and our system captures features of the spatial information in the region around the centroid of the drop for downstream applications.
% , our system leverages a CNN to extract spatial information across a patch around the centroid of the drop.

% We show in Fig.~\ref{fig:location}a our algorithm's detection accuracy at detecting the deposited drops, and the results show that the 25~\uL~is correctly detected 90--100\% across all locations. While for the 10~\uL~sample it is correctly detected 90--100\% only at the center and vertical locations, while being substantially lower at 30 and 50\% at the corner and horizontal location respectively. 

\noindent {\bf Macro-scale characterization.} Here, we characterize electrode sensitivity spatially across the tablet screen. These variations are often due to hardware non-idealities such as variations in dielectric thickness, grounding, and shielding around the sides of the screen compared to the center. Tackling this challenge is important to ensure that measurements of the same liquid and volume across different areas of the screen are uniform.

To characterize this, we performed an experiment by depositing a 50~\uL~drop of water at the center of each electrode cell at predefined intervals until the screen was covered. We repeated this procedure six times. In these six maps if a measurement occurred at the same pixel location across multiple maps, we averaged the results. In the end, we had a reading for 546 cells across the full grid of 52 $\times$ 32 = 1664 cells. 

Given calibration points $(x_i, y_i, s_i)$ where $s_i$ is the screen's response, we generate the interpolated heatmap $S(x,y)$ using a radial basis function with thin plate spline kernel using a smoothing factor $\lambda=3$. 

\begin{equation}
S(x, y) = \sum_{i=1}^{N} w_i \cdot \phi(\|(x,y) - (x_i,y_i)\|)
\end{equation}

\noindent where $N$ is the number of calibration points, $r = | (x, y) - (x_i, y_i) |$ is the Euclidean distance between the interpolation point $(x, y)$ and each calibration point $(x_i, y_i)$, and the basis function is defined as $\phi(r) = r^2 \log(r)$. The weights $w_i$ are determined by solving the regularized linear system:

\begin{equation}
\mathbf{A}\mathbf{w} = \mathbf{s} + \lambda\mathbf{w}
\end{equation}

\noindent with $A_{ij} = \phi(\| (x_i, y_i) - (x_j, y_j) \|)$, $w = [w_1, \dots, w_N]^T$, and $s = [s_1, \dots, s_N]^T$.

The sensitivity map in Fig.~\ref{fig:location}b, shows two clusters to the left and right of the screen which are sensitive, while the sides and the center have low sensitivity. 
% We show a histogram of the relative sensitivity, normalized from 0 to 1 in Fig.~\ref{fig:location}c, we note that the mean is 0.57.

The compensation factor in each grid cell is normalized using the function $compensation\_map = \frac{min(interpolated\_value)}{interpolated\_value + \epsilon}$. Here the minimum value across all calibration points corresponds to the most responsive region of the screen. This compensation map represents the inverse of the interpolated sensitivity, and $\epsilon$ is a small term to prevent numerical instability from division by zero. The compensation is subsequently applied by element-wise multiplying the measured heatmap with the computed compensation map.

\subsubsection{Temporal noise characterization and reduction}
\begin{figure*}
\centering
{\includegraphics[keepaspectratio, width=\linewidth]
    {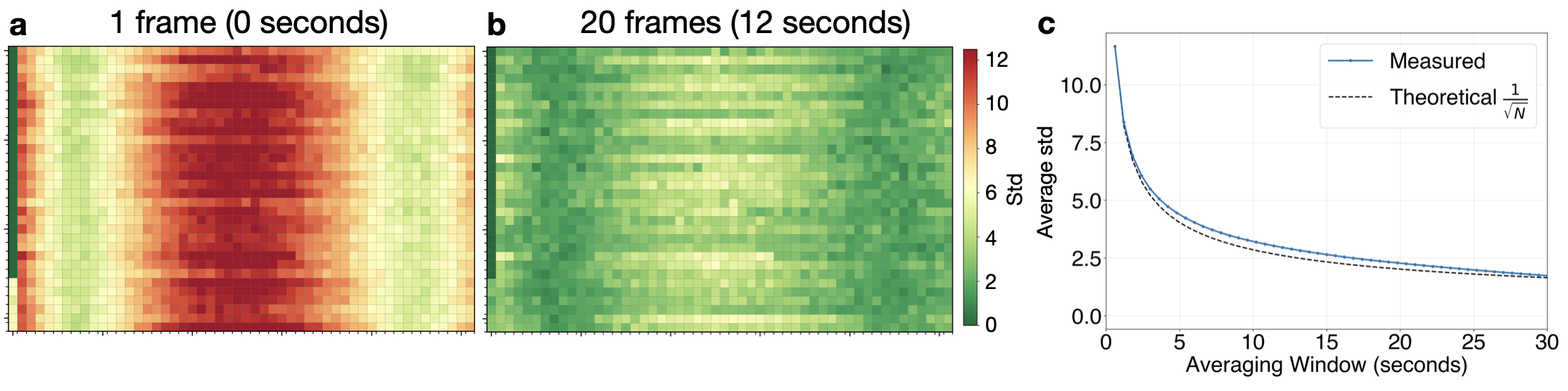}}
  \vspace{-2.5em}
  \caption{{\bf Noise characterization and reduction.} {\bf (a)} Standard deviation of capacitance readings across the screen for a single frame showing regions of higher and lower noise. {\bf (b)} Averaging over 20 frames (12~s) reduces noise. {\bf (c)} Average noise across screen decreases with longer averaging windows following the theoretical $\frac{1}{\sqrt{N}}$ trend.}
  \vspace{-1em}
  \label{fig:noise}
\end{figure*}

Here, we characterize the touchscreen's sensor noise over time. We note that because the touchscreen is primarily designed to detect touch locations rather than precise capacitance values, it is inherently not optimized for low-noise measurements. To characterize the noise profile, we record the tablet capacitance readings over the course of 30 seconds without anything on the screen. We show in Fig.~\ref{fig:noise}a the standard deviation per pixel across the full screen. We can observe the standard deviation is high in the center of the screen, and lower at the sides, the average standard deviation is 8. To reduce the effect of noise we perform averaging across each frame. Fig.~\ref{fig:noise}b shows the effect of averaging across 20 frames (12 seconds), with an average standard deviation across the screen of 2.8. The relationship between averaging time and the average standard deviation across the screen is shown in Fig.~\ref{fig:noise}c, the plot shows the theoretical relationship of $\frac{1}{\sqrt{N}}$, where $N$ is the number of frames being averaged over as well as the measured decrease in standard deviation over 30 seconds. The plot shows that the measured readings decrease with a trend closely following the theoretical model.

\red{\subsection{Data collection and protocol.}}
\label{sec:datacollection}

\begin{table}[h]
\centering

\begin{subtable}{\linewidth}
\centering
\begin{tabular}{lccc}
\toprule
Count & Unspiked & Spiked & Total \\
\midrule
\# drops    & 77 & 61  & 138 \\
\# sessions & 7  & 6   & 13  \\
\bottomrule
\end{tabular}
\caption{\red{Soda spiking}}
\end{subtable}
% \hfill

\begin{subtable}{\linewidth}
\centering
\begin{tabular}{lccc}
\toprule
Count & Unadulterated & Adulterated & Total \\
\midrule
\# drops    & 88 & 72 & 160 \\
\# sessions & 8  & 7  & 15  \\
\bottomrule
\end{tabular}
\caption{\red{Wine adulteration}}
\end{subtable}
% \hfill

\begin{subtable}{\linewidth}
\centering
\begin{tabular}{lcccc}
\toprule
Count & Unadulterated & Salt-adulterated & Detergent-adulterated & Total \\
\midrule
\# drops    & 44 & 33 & 41 & 118 \\
\# sessions & 4  & 3  & 4  & 11  \\
\bottomrule
\end{tabular}
\caption{\red{Milk adulteration}}
\end{subtable}

\caption{\red{\textbf{Dataset composition for adulteration detection tasks.} Dataset composition table for \textbf{(a)} soda spiking, \textbf{(b)} wine adulteration, \textbf{(c)} milk adulteration.}}
\label{tab:adulteration_data}
\end{table}

\begin{table}[h]
\centering
\begin{subtable}{\linewidth}
\centering
\begin{tabular}{lccc}
\toprule
Count & 0--20 ng/$\mu$L & 40--80 ng/$\mu$L & Total \\
\midrule
\# drops    & 124 & 165 & 289 \\
\# sessions & 12  & 16  & 28  \\
\bottomrule
\end{tabular}
\caption{\red{DNA concentration}}
\end{subtable}

\begin{subtable}{\linewidth}
\centering
\begin{tabular}{lccccccc}
\toprule
Count & 0\% & 10\% & 30\% & 50\% & 80\% & 100\% & Total \\
\midrule
\# drops    & 77 & 122 & 78 & 159 & 87 & 9 & 532 \\
\# sessions & 7  & 12  & 8  & 16  & 8  & 1 & 52  \\
\bottomrule
\end{tabular}
\caption{\red{Ethanol concentration}}
\end{subtable}

\begin{subtable}{\linewidth}
\centering
\begin{tabular}{lccc}
\toprule
Count & $5 \times 10^{-5}$--$10^{-4}$ M & $10^{-4}$--$10^{-1}$ M & Total \\
\midrule
\# drops    & 110 & 121 & 231 \\
\# sessions & 10  & 12  & 22  \\
\bottomrule
\end{tabular}
\caption{\red{NaCl concentration}}
\end{subtable}

\caption{\red{\textbf{Dataset composition for concentration detection tasks.} Dataset composition table for \textbf{(a)} DNA concentration classification, \textbf{(b)} ethanol concentration classification, \textbf{(c)} NaCl concentration classification.}}
\label{tab:concentration_data}
\end{table}

\begin{table}[h]
\centering
\begin{subtable}{\linewidth}
\centering
\begin{tabular}{lcccc}
\toprule
Count & Deionized water & Ethanol & Tap water & Total \\
\midrule
\# samples    & 20 & 16 & 18 & 54 \\
\# sessions & 5  & 4  & 5  & 14 \\
\bottomrule
\end{tabular}
\caption{\red{Basic liquids classification}}
\end{subtable}

\begin{subtable}{\linewidth}
\centering
\begin{tabular}{lccc}
\toprule
Count & Unspiked & Spiked & Total \\
\midrule
\# samples    & 16 & 16 & 32 \\
\# sessions & 4  & 4  & 8 \\
\bottomrule
\end{tabular}
\caption{\red{Soda adulteration detection}}

\end{subtable}

\medskip

\begin{subtable}{\linewidth}
\centering
\begin{tabular}{lccc}
\toprule
Count & 0.001\% & 0.01\% & Total \\
\midrule
\# samples    & 20 & 20 & 40 \\
\# sessions & 5  & 5  & 10 \\
\bottomrule
\end{tabular}
\caption{\red{NaCl concentration detection}}
\end{subtable}

\caption{\red{\textbf{Dataset composition for through-container sensing tasks.} Dataset composition table for \textbf{(a)} basic liquid classification, \textbf{(b)} soda adulteration detection, \textbf{(c)} NaCl concentration detection.}}
\label{tab:through_container_data}
\end{table}

\begin{table}[h]
\centering
\begin{tabular}{llccccccc}
\toprule
Split & Count & 0\% & 10\% & 30\% & 50\% & 80\% & 100\% & Total\\
\midrule
\multirow{2}{*}{Training (Day 1)}
& \# drops    & 77 & 61 & 39 & 79 & 40 & 9 & 305\\
& \# sessions & 7  & 6  & 4  & 8  & 4  & 1 & 30\\
\midrule
\multirow{2}{*}{Testing (Day 74)}
& \# drops    & 44 & 36 & 36 & 39 & 44 & 43 & 245\\
& \# sessions & 4  & 4  & 4  & 4  & 4  & 4 & 24\\
\bottomrule
\end{tabular}
\caption{\red{\textbf{Cross-day ethanol concentration dataset composition.}}}
\label{tab:crossday_data}
\end{table}

\red{We summarize here the data collection protocol used across all tasks in the paper, including the total number of unique liquid samples (drops), the number of data collection sessions and days, the cleaning procedures used to prevent cross-contamination, and the specifications of containers used for through-container sensing experiments. All datasets consist of physically distinct liquid drops; no synthetic data augmentation is used.}

\subsubsection{\red{Data collection sessions and spatial placement}}

\red{Data was collected across multiple sessions. A session is defined as a continuous data collection period from placing liquid drops or containers on the touchscreen to their removal. In all experiments, liquid drops and containers were placed randomly and uniformly across the touchscreen surface to avoid bias toward specific screen locations and to expose the model to spatial non-uniformities of the touchscreen.}

\red{The number of samples and sessions for each task is summarized in Tables~\ref{tab:adulteration_data}--~\ref{tab:through_container_data}. Table~\ref{tab:adulteration_data} summarizes the data composition for each adulteration detection task. Table~\ref{tab:concentration_data} summarizes the data composition for each adulteration detection task. Table~\ref{tab:through_container_data} summarizes the data composition for each adulteration detection task.}

\red{To evaluate robustness over extended time spans, we conduct a cross-day evaluation using the ethanol concentration classification task. The model is trained exclusively on data collected on one day and evaluated on data collected 74 days later, with no overlap in sessions or samples between training and testing. All samples in both training and testing are placed randomly across the touchscreen surface. Dataset composition for this evaluation is summarized in Table~\ref{tab:crossday_data}.}

{\subsubsection{\red{Cleaning and Cross-Contamination Prevention}}

\red{To prevent cross-contamination between samples, the touchscreen surface and any containers were cleaned between consecutive measurement sessions. After each session, the screen surface was wiped with paper towel wipes and isopropyl wipes, and allowed to dry completely before the next measurement. Containers used in through-container experiments were rinsed with deionized water and dried between samples.}

\subsection{Benchmarks}
\label{sec:app-bench}

\begin{figure}[h]
\centering
{\includegraphics[keepaspectratio, width=\linewidth]
    {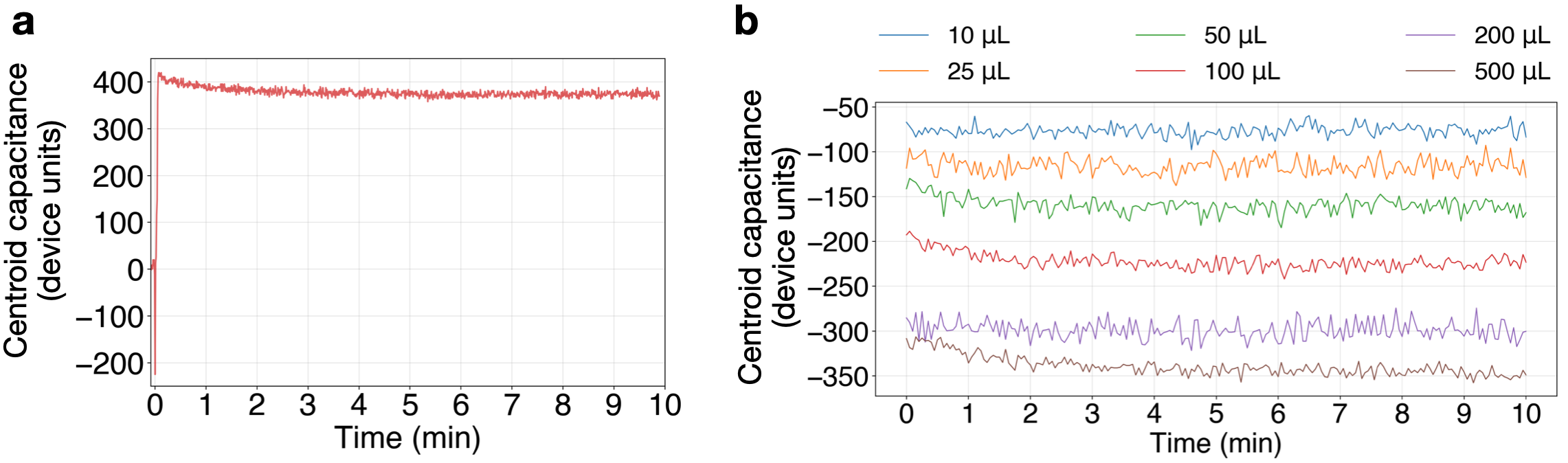}}
  \caption{{\bf Stability of capacitance measurements over time.} Capacitance for {\bf (a)} the priming drop and {\bf (b)} test droplets of different volumes. The plots shows that the capacitance readings remain stable over the course of 10 minutes.}
  \label{fig:lockzone}
\end{figure}

\noindent {\bf Measurement stability over time.} We evaluate the stability of capacitance readings for the priming drop and the test liquid samples on the tablet screen. We show in Fig.~\ref{fig:lockzone}a that the measured capacitance at the centroid of the priming drop remains stable across the measured period of ten minutes with a mean and standard deviation of 376 $\pm$ 7 device units. We also perform this measurement for water samples of different volumes across an 10 minute period and show in Fig.~\ref{fig:lockzone}b that these values remain stable across the measured period with standard deviations ranging from 6 to 10 device units across the measurement period. There is no substantial relationship between the standard deviation of the measurements with the volume of the liquid drops. We note that stability across these time scales is sufficient for averaging the effects of noise and for recording a measurement. 

\begin{table}
\centering
\begin{tabular}{lcc}
\toprule
 & Round-shaped cup & Heart-shaped cup \\
\midrule
\raisebox{3\height}{Container dimensions}
& \includegraphics[width=0.1\linewidth]{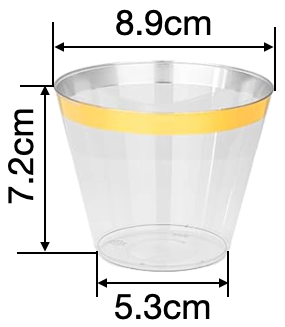}
& \includegraphics[width=0.1\linewidth]{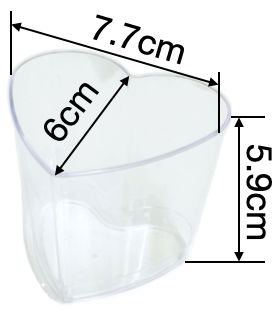} \\
Base thickness (mm) & 0.6 & 1.1 \\
\bottomrule
\end{tabular}%
\vspace{0.5em}
\caption{\red{Specifications of containers used in through-container sensing.}}
\label{tab:container_specs}
\vspace{-1em}
\end{table}

\begin{figure}[h]
\centering
\includegraphics[keepaspectratio, width=\linewidth]
    {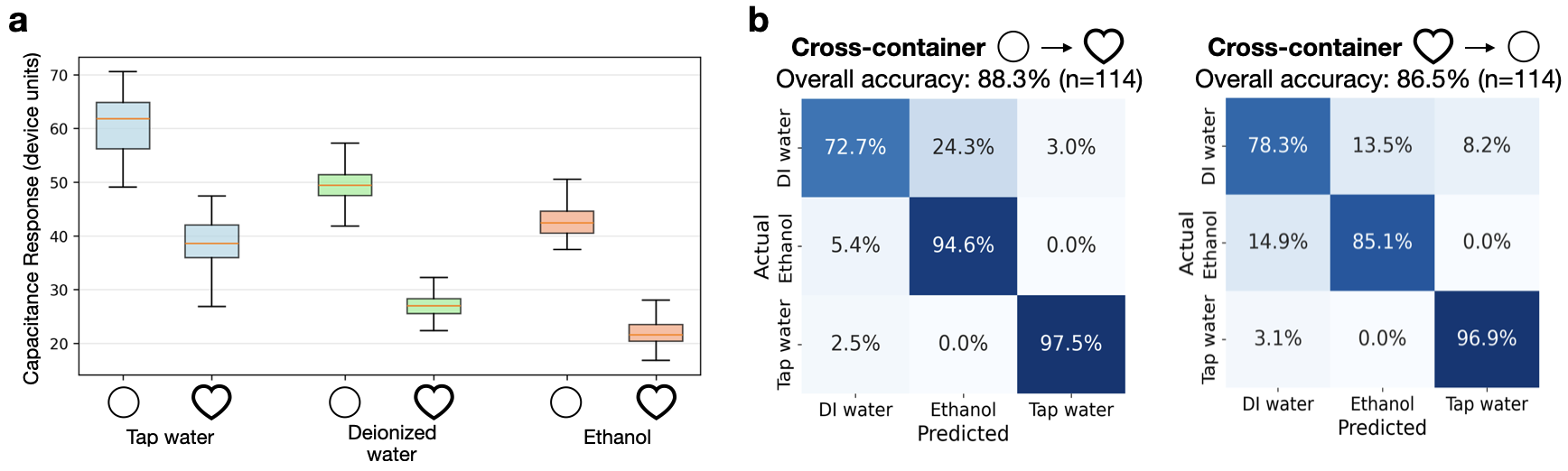}
  \caption{\red{{\bf Cross-container generalization with thickness compensation. } \textbf{(a)} Capacitance responses for round ($\bigcirc$) vs. heart ($\heartsuit$) shaped containers across three liquids. \textbf{(b)} Confusion matrix for cross-container classification: trained on round container, tested on heart container with thickness-based compensation.}}
  \vspace{-1em}
  \label{fig:cross-container}
\end{figure}

\noindent \red{{\bf Cross-container generalization. } To evaluate the robustness of our system across different container types, we conducted a benchmark study using two types of plastic containers with base geometries and base thicknesses.}

\red{We selected two containers (Table~\ref{tab:container_specs}): (1) Round plastic cup: cylindrical geometry, 0.6~mm base thickness; (2) Heart-shaped cup: non-circular geometry, 1.1~mm bottom thickness. Both containers feature thin, slightly concave bases that rest on the touchscreen surface with the outer rim making full contact.}

\red{As containers with different shaped bases have varying dimensions, direct comparison in the image space is challenging. To address this, we extract a single capacitance feature from the image of the container (mean of all positive pixels) to represent a sample. We show in Fig.~\ref{fig:cross-container}a that this feature is sufficiently predictive at distinguishing between different liquids.}

\red{The figure also shows that the base thickness has a direct effect on the attenuation of the capacitive signal. The heart-shaped container, with nearly twice the bottom thickness (1.1~mm vs 0.6~mm), exhibits lower capacitance values across all liquids. Importantly, the relative ordering of liquid classes remains consistent across containers. These results suggest that compensation strategy could enable cross-container generalization.}

\red{To compensate for thickness-induced signal attenuation, we derive a label-preserving transformation. Let $\mathbf{x}_{train}$ and $\mathbf{x}_{test}$ denote feature vectors from the training and testing containers, respectively, with thicknesses $t_{train}$ and $t_{test}$.}

\red{The compensation ratio is defined as:
\begin{equation}
\alpha = \frac{t_{test}}{t_{train}}
\end{equation}}

\red{Rather than uniformly scaling all features by $\alpha$ (which would artificially increase variance), we apply an additive offset that adjusts only the distribution's center:}

\red{\begin{equation}
\begin{alignedat}{2}
\Delta &= (\alpha - 1)\cdot \bar{\mathbf{x}}_{\text{test}} \\
\mathbf{x}_{\text{comp}} &= \mathbf{x}_{\text{test}} + \Delta
\end{alignedat}
\end{equation}}

\red{where $\bar{\mathbf{x}}_{test} = \frac{1}{N}\sum_{i=1}^{N} \mathbf{x}_{test}^{(i)}$ is the mean feature vector computed over all test samples. This formulation ensures that the compensated mean aligns with the training distribution, while the variance remains unchanged, maintaining class separability.}

\red{We trained a Random Forest classifier on one container type and evaluated it on the other. Without compensation, the model fails to generalize (33.3\% accuracy equals random chance). With thickness-based compensation, cross-container accuracy reaches 86.5 to 88.3\%, demonstrating effective domain adaptation. Fig.~\ref{fig:cross-container}b and Fig.~\ref{fig:cross-container}c shows the confusion matrix for the compensated cross-container classification.}

\red{These results demonstrate that our system can generalize to unseen container bases with different thicknesses and shapes. The shape of the container can be controlled for by distilling the capacitance image into a single scalar feature, while container thickness introduces a systematic bias in absolute signal magnitude, which can be corrected through simple physics-based compensation.}

% \begin{figure}
% \centering
% \includegraphics[keepaspectratio, width=.8\linewidth]
%     {figs/bm1.png}
%   \caption{{\bf Benchmarks} {\bf (a)} Water bath used to heat up liquid samples for experiment. {\bf (b)} Effect of temperature on liquid drops placed on touchscreen.}
%   \vspace{-1em}
%   \label{fig:bm1}
% \end{figure}

\begin{figure}[h]
\centering
\includegraphics[keepaspectratio, width=.8\linewidth]
    {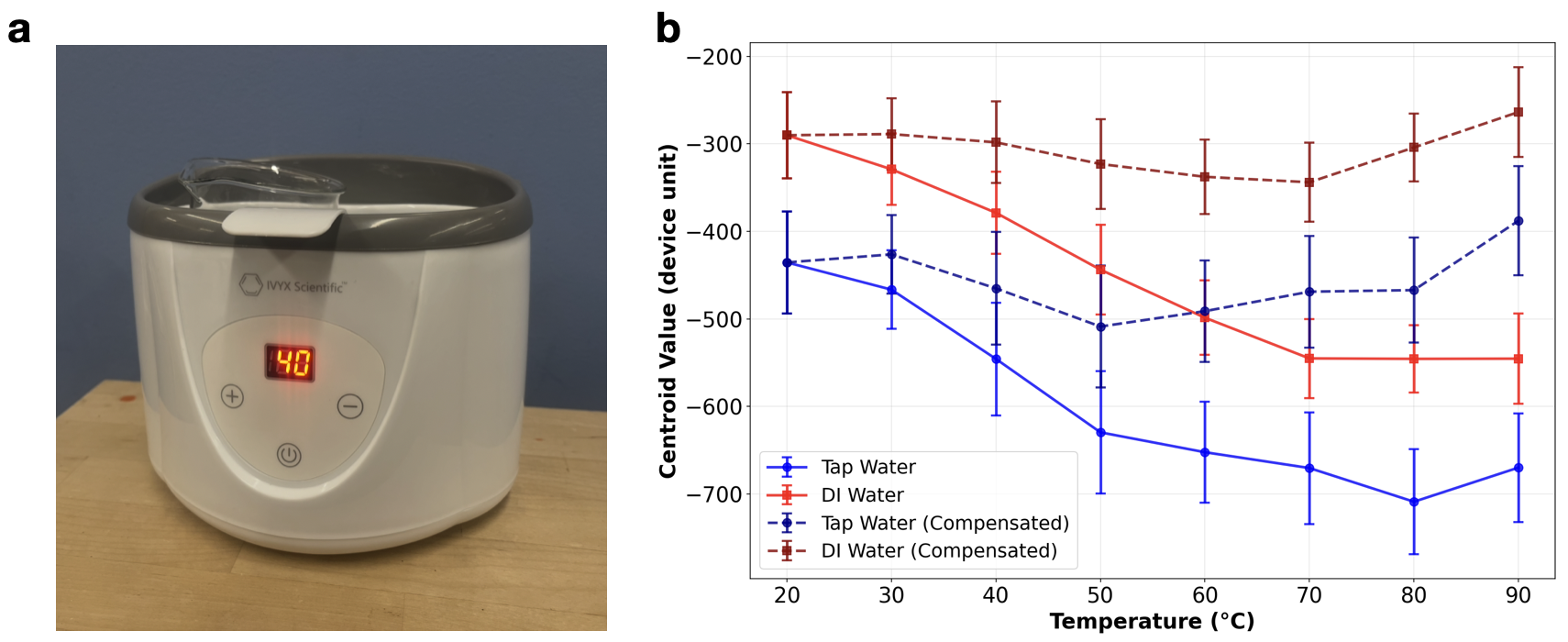}
  \caption{\red{{\bf Effect of temperature.} {\bf (a)} Water bath used to heat up liquid samples for experiment. {\bf (b)} Temperature compensation for cross-temperature classification.}}
  \vspace{-1em}
  \label{fig:tempcomp}
\end{figure}

\noindent \red{{\bf Effect of temperature.} We examine the effect of temperature on measured capacitance using tap water and deionized water. We heated up the liquid samples in a tabletop water bath~\cite{ivyx} from 20--90$^\circ$C (10$^\circ$C intervals). In Fig.~\ref{fig:tempcomp}b, the solid lines for tap water (blue) and deionized water (red) show that the measured capacitance increases for both liquids, which is due to conductivity increase as temperature increases.}

\red{Given the observed temperature effect, we investigate whether a compensation algorithm can maintain classification accuracy across temperatures. For each liquid, we split the data in half: the first half fits a linear model $y = At + b$ (where $t$ is temperature and $y$ is measured centroid capacitance), and the second half is used for evaluation.}

\red{The compensation offset at temperature $t$ is $\Delta(t) = (At + b) - (20A + b)$, normalizing all measurements to 20$^\circ$C. Fig.~\ref{fig:tempcomp}b shows the compensation effectively flattens the temperature-dependent drift for both liquids. To evaluate cross-temperature generalization, we trained a logistic regression classifier on centroid features at each temperature and tested on all other temperatures. The compensation achieves 86.1\% overall accuracy (69.7\% without compensation).}

\begin{figure}
\centering
{\includegraphics[keepaspectratio, width=\linewidth]
    {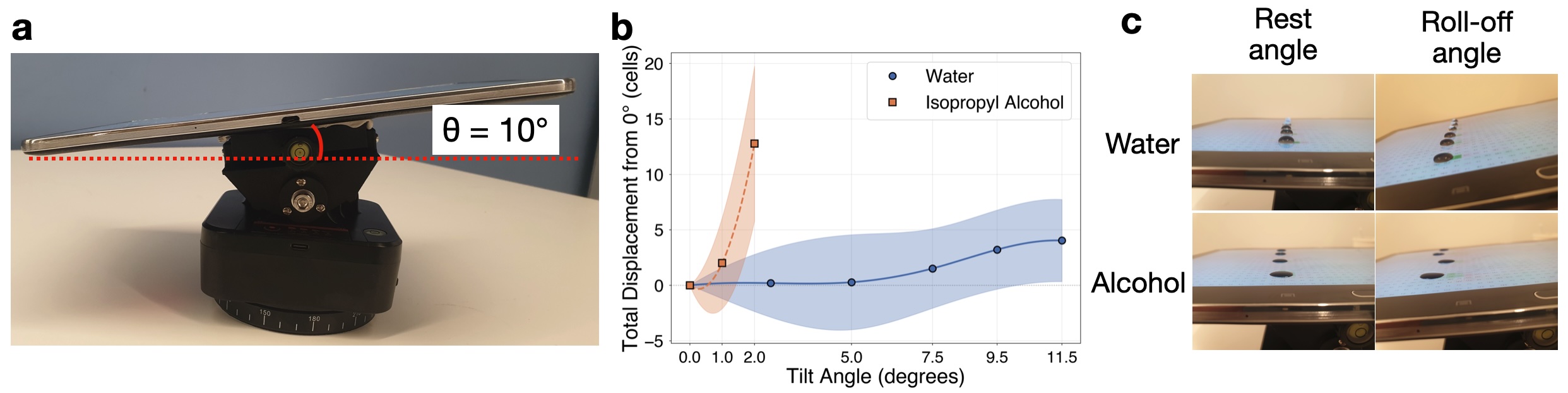}}
  \caption{{\bf Effect of tilting on liquids.} {\bf (a)} Motorized platform to control tablet tilt. {\bf (b)} Displacement of water and alcohol drops from their original position for different tilt angles. {\bf (c)} Side profile of water and alcohol when tablet is flat, and the roll-off angle when the drops start to slide across the surface.}
  \label{fig:tilt}
\end{figure}

\noindent {\bf Effect of device tilt.} \red{We conducted a benchmark study to evaluate the effect of device tilt on our system.} We placed the tablet on a motorized tilt platform~\cite{tilt} (Fig.~\ref{fig:tilt}a). We deposited ten drops of water and isopropyl alcohol (IPA) onto the surface and varied the tilt angle at increments of approximately 1 to 3$^{\circ}$. We show in Fig.~\ref{fig:tilt}b how the droplets displace from their original position as tilt angle increases. In the case of water, which has a surface tension of $\gamma=72.0~mN/m$, the droplets start moving at 5$^{\circ}$ and this displacement increases till its roll-off angle of 11.5$^{\circ}$ when it slides off the tablet. In contrast, IPA, which has a lower surface tension of $\gamma=21.7~mN/m$, has a much higher rate of displacement with a roll-off angle of 2$^{\circ}$. While water had a displacement of approximately 4 cells at the roll-off angle, alcohol had a displacement around three times higher at 12 cells. The behavior of the drops at the rest and roll-off angles are visualized in Fig.~\ref{fig:tilt}c.

\noindent \red{{\bf Physics-informed model.} We perform a benchmark study using physics-informed features that capture different aspects of the liquid's capacitive response and Random Forest classification. We show in Table~\ref{tab:physics_features} the list of selected features and their physical interpretation.}

\red{We evaluated the physics-informed model on three adulteration detection tasks. Fig.~\ref{fig:physics_results} shows the confusion matrices. The model achieves 84.5\% accuracy on soda spiking detection, 81.7\% on wine adulteration detection, and 95.6\% on milk adulteration detection. These results demonstrate that physics-informed features alone can achieve reasonable classification performance.}

\begin{table}[h]
\centering
\small
\begin{tabular}{p{2.7cm}p{9cm}>{\centering\arraybackslash}p{2.5cm}}
\toprule
\textbf{Feature} & \textbf{Physical Interpretation} & \textbf{Feature importance (\%)} \\
\midrule
Spatial spread & Weighted variance of sensor positions; measures the effective contact area influenced by surface tension & 25.2 \\
Skewness & Asymmetry of capacitance distribution; indicates non-uniform drop spreading or edge effects from wetting properties & 11.7 \\
Minimum value & Peak capacitance response; indicates the strongest electrical coupling point, influenced by liquid conductivity, permittivity and drop geometry & 9.6 \\
Centroid value & Capacitance at the weighted center of the drop; directly reflects permittivity and conductivity at the point of maximum coupling & 8.8 \\
Standard deviation & Variability in capacitance across the drop; relates to non-uniformity in drop shape and thickness from surface tension effects & 8.4 \\
Mean absolute value & Average magnitude of capacitance response; proportional to overall liquid permittivity and conductivity & 8.3 \\
25th percentile & Lower quartile capacitance; characterizes the distribution of strong coupling regions & 8.3 \\
Value range & Difference between maximum and minimum; related to the conductivity and permittivity of the liquid & 7.5 \\
75th percentile & Upper quartile capacitance; characterizes the distribution of weak coupling regions & 7.3 \\
Maximum value & Peak capacitance response; captures the fringing effect of the liquid drop & 5.1 \\
\bottomrule
\end{tabular}
\caption{\red{Physics-informed features and their relationship to liquid properties.}}
\label{tab:physics_features}
\end{table}

\begin{figure}
\centering
\includegraphics[keepaspectratio, width=.8\linewidth]
    {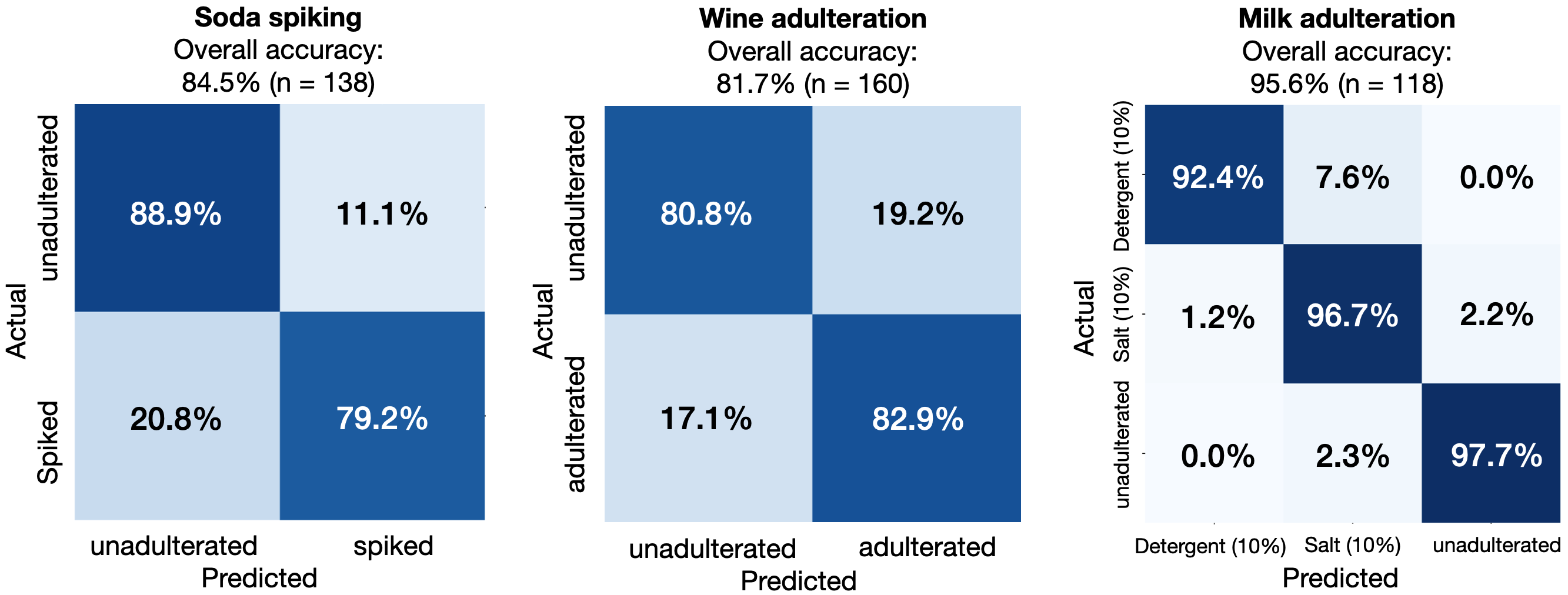}
\caption{\red{\textbf{Confusion matrices for adulteration detection using physics-informed Random Forest model.} \textbf{(a)} Soda spiking. \textbf{(b)} Wine adulteration. \textbf{(c)} Milk adulteration.}}
  \vspace{-1em}
  \label{fig:physics_results}
\end{figure}

\red{However, the accuracies are 5-10 percentage points lower than the CNN-based approaches, suggesting that while physics-informed features provide a strong foundation, the CNN can learn additional subtle patterns from the raw sensor data that are difficult to manually specify.
That being said, further feature engineering could be performed to develop a hybrid model combining physics-informed features, and data-driven features for use by a downstream classifier.}

\noindent \red{\textit{Feature importance analysis.} An important advantage of using physics-informed features is its interpretability. We perform a feature importance analysis which reveals that spatial spread (25.2\%) and skewness (11.7\%) are the most discriminative features, both of which primarily correspond to surface tension effects that affect drop spreading and wetting behavior. Following these, the minimum value (9.6\%) and centroid value (8.8\%) have the most discriminative value, and they are associated with the  liquid’s conductivity and permittivity.}

\begin{figure}[h]
\centering
{\includegraphics[keepaspectratio, width=\linewidth]
    {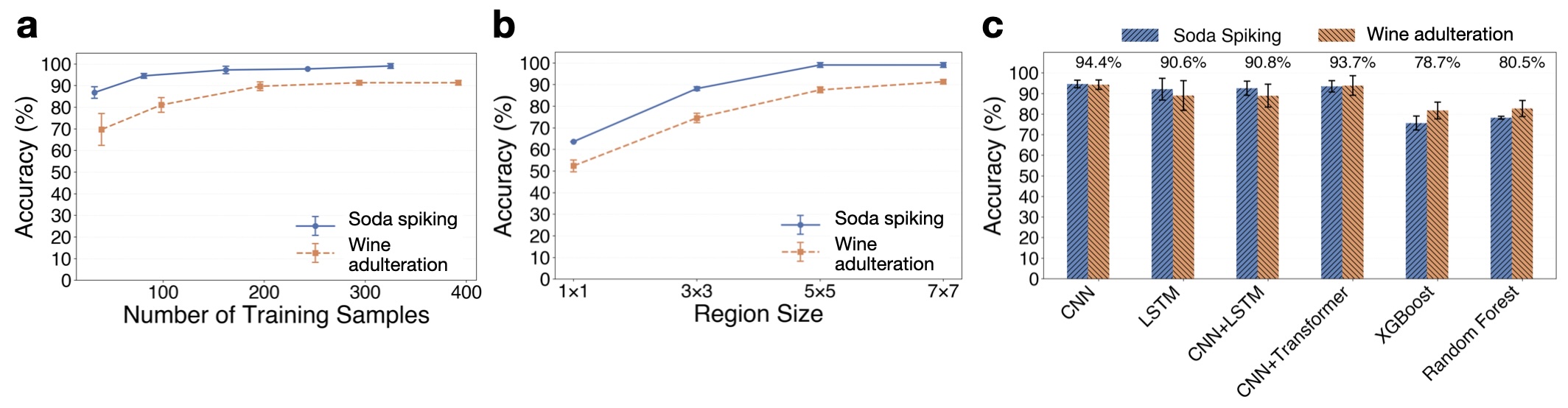}}
  \caption{{\bf Ablation study.} Effect of {\bf (a)} number of training samples, {\bf (b)} region size around centroid, {\bf (c)} machine learning model.}
  % \vspace{-1em}
  \label{fig:ablation}
\end{figure}

\noindent {\bf Ablation study.} We perform an ablation study on the liquid adulteration datasets to examine the effect of different system design choices on performance (Fig.~\ref{fig:ablation}). Specifically, we examine the effect of number of training samples, region size around the centroid, and machine learning model on system performance. For these experiments, we consider a subset of the adulteration classes: for soda spiking and wine adulteration we compared to ethanol of 10, 20, 30\% as a matter of computational efficiency to examine the overall effect of the system parameters. Here we used 500~\uL~drop sizes.

For the training dataset size (Fig.~\ref{fig:ablation}a) we evaluated the affect of changing the size of the training set from 0, 25, 50, 75, 100\% of the original size. We find that as expected this increases the accuracies with diminishing returns. In the soda and wine dataset, we find that there are no substantial performance increases beyond approximately 200 training samples.

For the region size (Fig.~\ref{fig:ablation}b), we find that accuracy increases steadily as the region size increases from a \dimens{1}{1} patch to a \dimens{7}{7} patch. In the \dimens{1}{1} case we use a fully-connected network instead of a CNN. For larger classes we use an adaptive CNN where for \dimens{2}{2} and \dimens{3}{3} classes there are only two convolutional layers without any pooling, while for larger patch sizes of \dimens{4}{4} to \dimens{6}{6} there is a single max pooling layer, and for \dimens{7}{7} there are 2 pooling layers. This result is expected as the screen area covered by a drop can expand to affect the electric field of a \dimens{7}{7} grid, and a larger grid captures this information.

Finally, we examined the effect of model architecture on system performance (Fig.~\ref{fig:ablation}c). We find that using a CNN only, LSTM only, CNN+LSTM and CNN+Transformer provide performance within a similar range of 91 to 94\%. Tree-based models such as XGBoost and Random Forest provide lowered accuracies of 79 to 81\%. The results suggest that global features about the region, can be captured by the tree-based models. However, deep learning modules better capture the non-linear relationship (Sec.~\ref{sec:theory}) between classes, with the CNN and LSTM producing higher accuracies. As CNN alone and LSTM alone are able to produce high accuracies, it suggests that non-linear feature combinations are sufficient to perform classification.

\begin{figure}
\centering
\includegraphics[keepaspectratio, width=.7\linewidth]
    {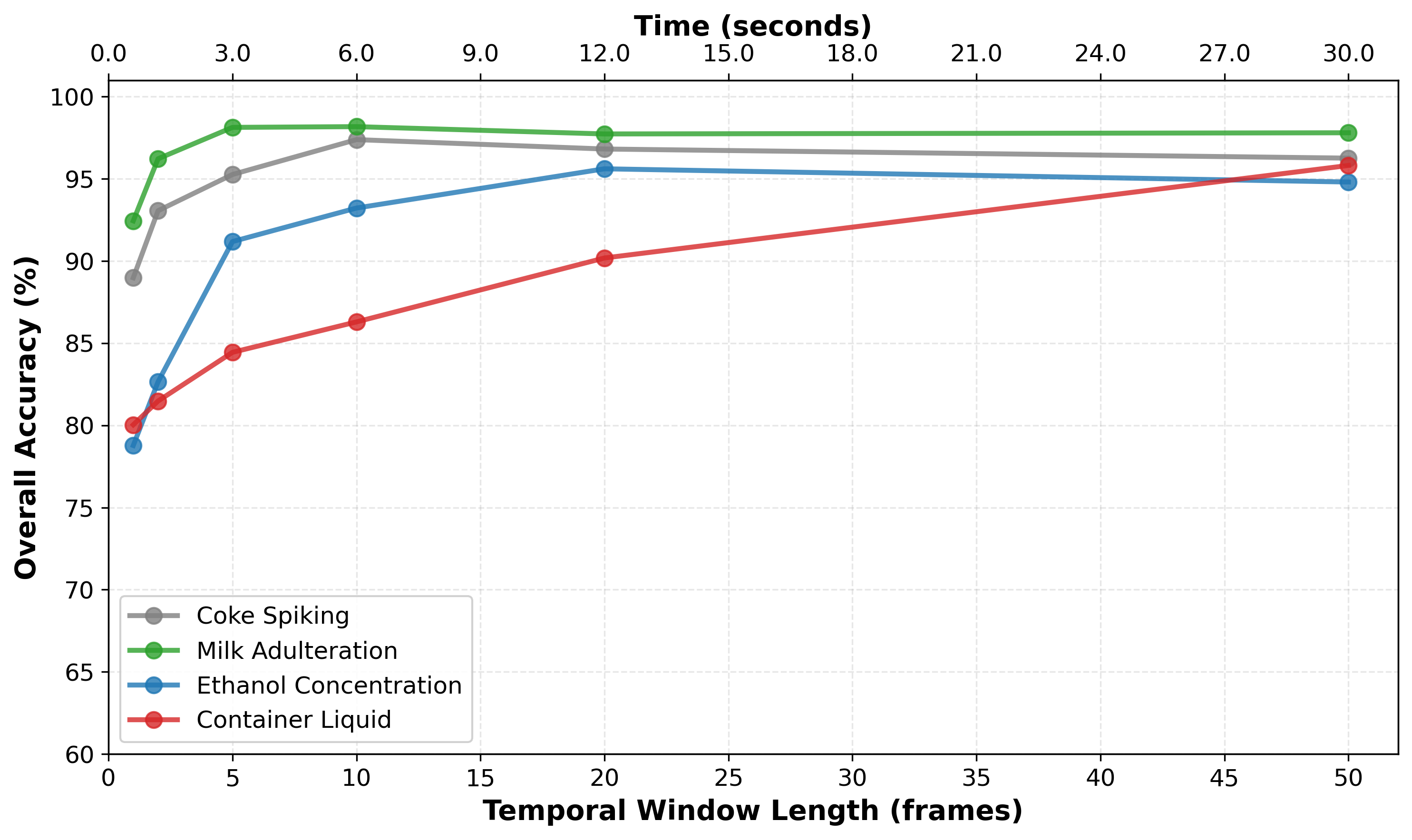}
  \caption{\red{{\bf Temporal length sensitivity analysis.} Effect of data's temporal length on classification task accuracy}}
  \vspace{-1em}
  \label{fig:temporal}
\end{figure}

\noindent \red{{\bf Temporal length sensitivity analysis. } We evaluate the effect of temporal context (number of training frames) on system performance for different liquid sensing tasks. We trained classifiers using sequences of 1, 2, 5, 10, 20, and 50 frames, with all frames provided as independent training samples. At test time, each frame is classified independently. As each frame corresponds to a 0.6-second interval, our deployed system achieves an inference latency of 0.6 seconds.}

\red{Fig.~\ref{fig:temporal} shows classification accuracy as a function of training temporal window length across all tasks. While we used 50-frame sequences (30 seconds) in our primary experiments to maximize accuracy, the results demonstrate that substantially shorter training sequences suffice for most applications.}
 
\red{Specifically, adulteration detection tasks (coke spiking, milk adulteration) achieve >95\% accuracy when trained on short sequences of 5 frames (3 seconds), and concentration detection of ethanol maintains >90\% accuracy at this temporal resolution. The through-container task, which involves sensing through opaque plastic barriers, requires longer training sequences, reaching 90\% accuracy when trained on 20-frame sequences (12 seconds).}

\red{Our design classifies each frame independently instead of averaging measurements over time windows, which provides three key advantages: 
\squishlist
\item \textit{Low inference latency}: Our models can classify liquids in real-time with 0.67-second latency per frame, enabling immediate feedback.
\item \textit{Data augmentation}: Training on individual frames rather than averaged sequences substantially increases the training dataset size. 
\item \textit{Noise robustness}: By training on noisy individual frames rather than averages, our model learns to establish decision boundaries that are robust to measurement noise. This produces classifiers that can perform inference on single, noisy measurements.
\squishend}

\end{document}